\documentclass[reprint,showpacs,preprintnumbers,nofootinbib,nobibnotes,aip,jcp,amsmath,amssymb,]{revtex4-1}
\usepackage{graphicx}
\usepackage{amsmath}	% required for `\align' (yatex added)
\usepackage{colordvi}
\newcommand{\upd}{\mathrm{d}}

\begin{document}

%\preprint{APS/123-QED}

\title{Differential capacitance of the electric double layer:
\\ The interplay between ion finite size and dielectric decrement}
% Force line breaks with \\
%\thanks{A footnote to the article title}%

\author{Yasuya Nakayama}
 \email{nakayama@chem-eng.kyushu-u.ac.jp}
\affiliation{Department of Chemical Engineering, Kyushu University,
Nishi-ku, Fukuoka 819-0395, Japan}

\author{David Andelman}
 \email{andelman@post.tau.ac.il}
\affiliation{
Raymond and Beverly Sackler School of Physics and Astronomy,
Tel Aviv University, Ramat Aviv 69978, Tel Aviv, Israel
}%

\date{\today}% It is always \today, today,

%%%%%%%%%%%%%%%%%%%%%%%%%%%%%%%%%%%%%%
\begin{abstract}
%%%%%%%%%%%%%%%%%%%%%%%%%%%%%%%%%%%%%%
We study the electric double layer by combining the effects
of ion finite size and dielectric decrement.
At high surface potential, both mechanisms can cause saturation of the
counter-ion concentration near a charged surface.
The modified Grahame equation and differential capacitance
are derived analytically for a general expression of a permittivity
$\varepsilon(n)$ that depends
on the local ion concentration, $n$, and
under the assumption that the co-ions are fully depleted from the surface.
The concentration at counter-ion saturation is found for any
$\varepsilon(n)$, and a criterion predicting which of the two mechanisms
 (steric vs. dielectric decrement)
is the dominant one is obtained.
At low salinity, the differential capacitance as function of surface
potential has two peaks (so-called camel-shape).
Each of these two peaks is connected to a saturation of counter-ion
 concentration caused either by dielectric decrement or by their finite size.
Because these effects depend mainly on the counter-ion concentration at the surface proximity,
for opposite surface-potential
polarity either the cations or anions play the role of counter-ions,
resulting in an asymmetric camel-shape.
At high salinity, we obtain and analyze the crossover in the differential capacitance
from a double-peak shape to a uni-modal one.
Finally,  several nonlinear models of
the permittivity decrement are considered, and we predict that the
 concentration at dielectrophoretic saturation shifts to higher concentration
than those obtained by the linear decrement model.
\end{abstract}

%\pacs{61.20.Qg, 82.47.Uv, 82.60.Lf}% PACS, the Physics and Astronomy
                             % Classification Scheme.
%\keywords{Suggested keywords}%Use showkeys class option if keyword
                              %display desired
%
%61.20.Qg Structure of associated liquids: electrolytes, molten salts, etc.
%82.47.Uv Electrochemical capacitors; supercapacitors
%82.60.Lf Thermodynamics of solutions
%82.70.Dd Colloids
%82.45.Gj Electrolytes
%78.30.cd Solutions and ionic liquids

\maketitle

%\tableofcontents

%%%%%%%%%%%%%%%%%%%%%%%%%%%%%%%%%%%%%%%%%%%%%%%%%
\section{Introduction}
%%%%%%%%%%%%%%%%%%%%%%%%%%%%%%%%%%%%%%%%%%%%%%%%%

The behavior of ions in liquids, or more specifically, in aqueous solutions near charged
surfaces, largely contributes  to our understanding of colloidal
interactions~\cite{2006Soft,Israelachvili2011Intermolecular},
transport in nano- and
micro-pores~\cite{Stone2004ENGINEERING,Sparreboom2009Principles}, and
electrochemical processes~\cite{Jayalakshmi2008Simple}.
Traditionally, the ionic profiles have been calculated using the
Poisson--Boltzmann~(PB)
model~\cite{2006Soft,Israelachvili2011Intermolecular}, which
includes electrostatic interactions and entropy of mobile ions dispersed
in a continuum medium of uniform permittivity, $\varepsilon$.
The PB model captures the formation of the so-called electric double layer~(EDL)
close to a charged surface, where the
counter-ions are loosely associated with the surface.
At small surface-charge densities, the PB model
is quite successful in explaining qualitatively many experimental
results. However,
at high surface-charge densities and for multivalent counter-ions,
the PB model fails to describe
the EDL even on a qualitative level~\cite{Bazant2009Towards}, and
other theories that take into account charge correlations and
fluctuations, have been
proposed~\cite{Henderson1979Application,Nielaba1985Packing,Netz2000Beyond,Moreira2000Strongcoupling,Abrashkin2007Dipolar,levy12:_dielec_const_of_ionic_solut,Levy2013Dipolar}.

Experiments on differential capacitance at low salt concentrations
indicate that the EDL width is a decreasing function of the surface
potential at low surface charge,  in accord with the PB predictions. However, the PB does not
predict correctly the increase in
the EDL width with the surface potential at higher surface
charge~\cite{Clavilier1977Etude,Valette1981Double,Valette1982Double}.
One attempt to explain this qualitative non-PB change of the EDL width was to
include  steric interactions between the (solvated)
ions~\cite{Bikerman1942XXXIX,KraljIglic1994Influence,Borukhov1997Steric,Kornyshev2007DoubleLayer,Kilic2007Steric_i,Biesheuvel2007Counterion,Dreyer2013Overcoming,Dreyer2014Mixture}
in the PB model.
In such a \textit{sterically-modified PB}~(smPB) model, the
saturation of the counter-ion build-up at the surface reaches closed-packing
density at high potentials. The model further predicts an increase of the EDL width with
the surface charge (or equivalently with surface potential), at high
surface charge conditions, in qualitative accord with experiments.

A second source of counter-ion saturation is caused by the dielectric decrement
characteristic to ionic solutions, and is due
to the effective polarizability of hydrated ions~\cite{Bikerman1942XXXIX,Sparnaay1972Electrical,Biesheuvel2005Volume,BenYaakov2011Dielectric,Hatlo2012Electric,LopezGarcia2013Influence,LopezGarcia2014Influence}.
This effective polarizability is related to the presence of a hydration shell
around ions in addition to the dielectric hole created by the ions
themselves.
Since ions usually have smaller dielectric constant than water,
inclusion of an ion in water creates a dielectric hole that reduces
the dielectric constant of the solution. However, for small and simple ions such as halides, this effect
does not have a substantial contribution to the net dielectric decrement.
A more significant effect is due to the hydration shell formed by water molecules
in the immediate proximity to an ion. In this layer, the polar water molecules are largely oriented along the
electrostatic field created by the ion, reducing the
orientational polarizability and leading to
a rather pronounced dielectric decrement.

The effective polarizability of hydrated ions in relation with
the EDL capacitance was first analyzed
by Bikerman~\cite{Bikerman1942XXXIX}, who predicted a shift in the
electrocapillary maximum due to the asymmetric effective
polarizabilities
between the cations and anions.
Depletion of a polyelectrolyte caused by dielectric decrement near the
surface was pointed out by Biesheuvel~\cite{Biesheuvel2005Volume}, and
the broadening of EDL by the dielectric decrement in multivalent ions and
mixed electrolytes was reported in
Refs.~\onlinecite{LopezGarcia2013Influence,LopezGarcia2014Influence}.

The saturation of counter-ion concentration that originates from the reduced
permittivity close to the surface is called {\it dielectrophoretic
saturation}, and was analyzed for solutions containing only
counter-ions that balance the surface charges~\cite{BenYaakov2011Dielectric}, as well as when
salt was added to the solution~\cite{Hatlo2012Electric}.
First, it was shown that the dielectric decrement can cause a
counter-ion saturation at the surface even without including the effect of
ion finite size~\cite{BenYaakov2011Dielectric}.
Later, it was pointed out that when the dielectric decrement is large, the
dielectrophoretic saturation will be the dominant effect even when the ion
finite size is taken into account.
Furthermore, the non-monotonic variation of the differential capacitance
with the surface charge can be ascribed not only to the ion finite size
as predicted by the smPB model, but also to the
dielectric decrement~\cite{Hatlo2012Electric}.

Since the counter-ion saturation at high surface charge can result
from two different origins, the following two points should be
elucidated. (i) Finding a general criterion to determine the dominant
mechanism (steric or dielectrophoretic) for the counter-ion saturation for general nonlinear
permittivity decrement; (ii) Combining the effects of dielectric decrement
and ion finite size in order to analyze the differential capacitance, not only
for the case of dielectrophoretic counter-ion saturation, but also for the
sterically-dominant one.

Another issue that needs further attention is the high salt regime.  As
the salt concentration increases, the range of the surface potential
where the PB model can be applied becomes smaller, and the counter-ion
concentration reaches a saturation value even at rather low values of the
surface potential. Using the
smPB model~\cite{Kornyshev2007DoubleLayer},  a uni-modal
(or bell-shape) differential capacitance was predicted at high salt concentration
above a threshold, $n_{b}>1/(8a^{3})$, where $n_{b}$ is the salt
concentration and $a$ is the lattice size, which roughly corresponds to the solvated ion
size~\cite{comment_Kornyshev}.
We note that a similar bell-shape differential capacitance has been
observed experimentally in ionic liquids~\cite{Islam2008Electrical},
while another type of bell-shape capacitance was predicted at an electrified
oil/water interface due to large organic
ions~\cite{GuerreroGarcia2012Inversion,GuerreroGarcia2013Enhancing}.
However, the additional effect of dielectric decrement on the emergence of
the bell-shaped differential capacitance has not yet been explored. For
ionic solutions, the dielectric decrement at high salt concentrations can
have a strong effect and, hence, change the camel-shape to bell-shape
crossover of the differential capacitance.

In this paper, we study the ionic behavior of an aqueous electrolyte
solution in the proximity of a surface having a constant charge
density or an externally controlled surface potential.
Our treatment is based on mean-field theory that includes both steric
ionic effects and dielectric decrement (dielectrophoretic).
The model is presented in Sec.~\ref{sec:model}, while in Sec.~\ref{sec:saturations},
a criterion for the dominant saturation of counter-ion concentration is found to be ionic specific.
We also present an expression for the modified Grahame equation in Sec.~\ref{sec:saturations},
and for the differential capacitance in Sec.~\ref{sec:difc}, as applicable
for general $\varepsilon(n)$ and finite ionic size.
We find that the EDL width exhibits a non-monotonic variation with the
surface charge, both for steric saturation and dielectrophoretic one, as is presented in
Secs.~\ref{sec:dielectrophoretic_saturation} and
\ref{sec:steric_saturation}.  In Sec.~\ref{sec:high_salinity}, based on
analytic results and numerical solutions, we explore the combined effect
of dielectric decrement and ion finite size on the differential
capacitance behavior, ranging from a camel-shape at low salt
concentrations to bell-shape at high salt concentrations.  Furthermore,
corrections due to nonlinear permittivity decrement are examined in
Sec.~\ref{sec:nonlinear_decrement}, because they can be rather
substantial at high salinity.

%%%%%%%%%%%%%%%%%%%%%%%%%%%%%%%%%%%%%%%%%%%%%%%%%%%%%%%%%%%%%%
\section{Model}
\label{sec:model}
%%%%%%%%%%%%%%%%%%%%%%%%%%%%%%%%%%%%%%%%%%%%%%%%%%%%%%%%%%%%%%

We consider a monovalent 1:1 electrolyte solution with bulk
salt concentration $n_{b}=n_{b}^{+}=n_{b}^{-}$. The aqueous solution occupies the $z>0$ region,
and is in contact with a planar surface located at $z=0$. This plane
has a constant charge density (per unit area), $\sigma$, or, equivalently
is held at a constant surface potential, $\psi_s$.
The mean-field free energy is expressed in terms of the ion
number densities (per unit volume), $n_{\pm}(z)$, and electrostatic potential
$\psi(z)$, and reads~\cite{BenYaakov2011Dielectric,Borukhov1997Steric}
\begin{align}
 F[n_\pm,\psi] &=\int_{0}^{\infty}\upd z
\left[-\frac{1}{2}\varepsilon_{0}\varepsilon({n_\pm})
\left|\psi'\right|^{2} +e(n_{+}-n_{-})\psi\right]
\nonumber\\
&+k_{B}T\int_{0}^{\infty}\upd z~
\big[n_{+}\ln(n_{+}a^{3}) + n_{-}\ln(n_{-}a^{3})\big]
\nonumber
\\
& +\frac{k_{B}T}{a^3} \int_{0}^{\infty} \upd z~ {\phi_w}\ln\phi_w,
\label{eq:free_energy}
\end{align}
where $\psi^\prime={\rm d}\psi/{\rm d}z$, $e$ is the electronic charge, $\varepsilon_{0}$ is the vacuum permittivity, $\varepsilon(n_\pm)$ is
the relative permittivity that depends in our model on the local ion density
$n_\pm(z)$, and, hence, implicitly on the distance $z$
from the surface,
$\phi_w(z)=1-a^{3}\sum_{\alpha=\pm}n_{\alpha}$
is the solute (water) volume fraction, $a$ is the lattice spacing that is roughly
equal to the solvated ion diameter,
$k_{B}$ is the Boltzmann constant, and $T$ is
the temperature.

The Euler--Lagrange equations determining the electrostatic potential and ion concentrations
are obtained by taking the variation with respect to $n_{\pm}$ and $\psi$ of the excess free energy defined from
Eq.~(\ref{eq:free_energy}) to be $\Delta F=F[n_\pm,\psi]-F[n_b,0]$.
These  equations are written as:
%
%%%%%%%%%%%%%%%%
\begin{align}
 \ln\frac{n_{b}}{\phi_{b}}
&=
 \ln\frac{n_{\pm}}{\phi_{w}}
\pm \frac{e\psi}{k_{B}T}
-\frac{\varepsilon_{0}}{2k_{B}T}\frac{\partial \varepsilon}{\partial
 n_{\pm}}
\left|\psi'\right|^{2},
\label{eq:electrochemical_potential}
\\
&\frac{\upd}{\upd z}\Big(\varepsilon_{0}\varepsilon(n_\pm(z))
\psi'\Big)= -e(n_{+}-n_{-})\, ,
\end{align}
where the volume fraction of bulk water is
\begin{equation}
 \phi_b=1-2a^3n_b \,.
\end{equation}
From the electro-chemical potential equality of
Eq.~(\ref{eq:electrochemical_potential}), the ion concentrations can be expressed as
%%%%%%%%%%%%%%%%
\begin{align}
 n_{\pm}(z) &= \frac{\rho_{\pm}}{\phi_b+a^3(\rho_{+}+\rho_{-})}\, ,
\label{eq:mpb_distribution}
\\
\rho_{\pm}(z) &=n_{b}\exp\left(\mp \frac{e\psi}{k_{B}T}
+\frac{\varepsilon_{0}}{2k_{B}T}
\frac{\partial \varepsilon}{\partial n_{\pm}} \left|\psi'\right|^{2}\right)\, ,
\label{eq:ppb_distribution}
\end{align}
where $\rho_\pm(z)$ is another ionic profile defined to
make the above expressions look simpler.
%%%%%%%%%%%%%%%%%%%%%%%%
%
The boundary condition at the $z=0$ surface is obtained from the charge
neutrality condition:
\begin{align}
-\varepsilon_{0}\varepsilon_{s}\psi'_s
&= \sigma\, ,
\label{eq:bc}
\end{align}
where $\sigma$ is the surface charge density,
$\varepsilon_{s}=\varepsilon(n_\pm(0))$ is the extrapolated value of the
permittivity at the $z=0$ surface, and $\psi'_s=\upd \psi/\upd z|_s$ is also evaluated at the same surface.
Moreover,  $\psi'=0$ is imposed at $z\to\infty$ (no electric field at
infinity).

One of the well-known attempts to go beyond the PB treatment was to
consider a surface layer called the Stern layer~\cite{Stern1924Zur} with reduced dielectric
constant and also to account for
specific adsorption of ions on the surface due to non-electrostatic
ion-surface interactions.
The Stern proximity layer was introduced to explain some experimental
observations which cannot be explained by the conventional PB
model, that are
magnitude of the capacitance
~\cite{Grahame1954Differential,Bonthuis2013Beyond} and its non-monotonic \(\psi_{s}\)
dependence~\cite{Grahame1954Differential}.
For precise
description of the Stern layer, the layer width and the dielectric
profile should be determined by explicit modeling of the molecular
interactions between ions, solvent and the surface.

We note that
such a successful attempt has been reported in Ref.~\onlinecite{Bonthuis2013Beyond}.
The Stern layer parameters extracted from
molecular dynamic simulations were combined with a PB model and resulted in
a quantitative agreement with experiments of the differential
capacitance at vanishing surface potential. In contrast to this, non-monotonic
\(\psi_{s}\)-dependence of experimental differential capacitance is
supposed to be largely due to the ionic profile in the diffuse layer.
\cite{Bikerman1942XXXIX,Kornyshev2007DoubleLayer,Kilic2007Steric_i,Bazant2009Towards,Hatlo2012Electric}

In our above described model, the Stern layer is not taken into
accounted, although, in principle, it can be incorporated by introducing
a layer with given thickness and low dielectric constant adjacent to the
surface.  We leave this refinement to future studies because the main
thrust of the present work is to study the interplay between finite
ionic size and dielectric decrement, and their effect on ionic profiles
and differential capacitance, especially dependence on \(n_{b}\) and \(\psi_{s}\).

It is experimentally known~\cite{Hasted1948Dielectric} that the relative permittivity of a bulk electrolyte
solution decreases linearly with salt concentration, at low salinity,
$n_{b}\lesssim 2$\,M.
This dependence
is written as $\varepsilon = \varepsilon_{w}-\gamma_{b}n_{b}$, where $\varepsilon_{w}$ is
the relative permittivity of the solvent (water) and $\gamma_{b}$ is a
coefficient measured in units of M$^{-1}$.
At higher $n_b$ values, however, the dielectric decrement shows
a more complex nonlinear dependence~\cite{levy12:_dielec_const_of_ionic_solut,Levy2013Dipolar}, which levels off
and has a smaller decrement than the linear one.

As we are interested in the EDL behavior where
the distribution of counter-ions and co-ions is highly inhomogeneous, we will explicitly take
into account the local variation of $\varepsilon(n_\pm)$ with $n_\pm(z)$.
Since no direct experiment reported so far the local variation of $\varepsilon(n_\pm)$ inside the EDL,
we will first use a {\it linear decrement model}, which linearly superimpose the separate contributions of each ionic species
\cite{BenYaakov2011Dielectric,Hatlo2012Electric,nonadditivity}
\begin{align}
\varepsilon(n_\pm) &=
 \varepsilon_{w}-\gamma_{+}n_{+}(z)-\gamma_{-}n_{-}(z)\, ,
\label{eq:linear_decrement}
\end{align}
where
$\gamma_{\pm}$ is the coefficient of the $\pm$ ion, respectively.
These coefficients are not known but can be determined by assuming
a simple relationship $\gamma_{b}=\gamma_{+}+\gamma_{-}$
and using the experimentally known values of $\gamma_b$ as in Ref.~\onlinecite{Hasted1948Dielectric}.
Such values of $\gamma_{\pm}$ are summarized in
Table~\ref{tbl:parameters} for several monovalent
cations and anions~\cite{Hasted1948Dielectric,Hatlo2012Electric}.

In the numerical calculations presented below
(within the linear decrement model of Eq.~(\ref{eq:linear_decrement})),
we will often use as an example the parameters of
a monovalent KCl aqueous solution with
$\gamma_{\rm{K}^+}=8\,\text{M}^{-1}$, $a_{\rm{K}^+}=0.662$\,nm for
K$^{+}$, and $\gamma_{\rm{Cl}^-}=3\,\text{M}^{-1}$, $a_{\rm
{Cl}^-}=0.664$\,nm for Cl$^{-}$ (see Table~I). In addition, the water
dielectric constant is $\varepsilon_{w}=78.3$ at $T=25^{\circ}$C.

%%%%%%%%%%%%%%%%%%%%%%%%%%%%%%%%%%%%%%%%%%%%%%%%%%%%%%%%%%%%%%%%%%%%%%%%%%%%%%%%%%%%%%%%%%%
\begin{table}[htbp]
\caption{\label{tbl:parameters}
Experimentally obtained coefficients of dielectric decrement~\cite{Hasted1948Dielectric}, $\gamma$, and
hydration diameter~\cite{Nightingale1959Phenomenological}, $a$,
for several monovalent cations and anions.
The ratio $d/a$ serves as our criterion to determine the mechanism of
 counter-ion saturation
 for the linear permittivity decrement (see text).
 The error bars of
 $\gamma$ were reported in Ref.~\onlinecite{Hasted1948Dielectric} to be  about $\pm 1$\,M$^{-1}$.
}
\begin{ruledtabular}
 \begin{tabular}{cccc|cccc}
   & $\gamma~[\text{M}^{-1}]$ & $a$\,[nm] & $d/a$
&
   & $\gamma~[\text{M}^{-1}]$ & $a$\,[nm] & $d/a$\\
  \hline
H$^{+}$  & 17 & 0.564 & 1.59
&
F$^{-}$  & 5 & 0.704 & 0.85 \\
Li$^{+}$  & 11 & 0.764 & 1.02
&
Cl$^{-}$  & 3 & 0.664 & 0.76 \\
Na$^{+}$  & 8 & 0.716 & 0.97
&
I$^{-}$  & 7 & 0.662 & 1.01 \\
K$^{+}$  & 8 & 0.662 & 1.05
&
OH$^{-}$  & 13 & 0.6 & 1.37 \\
Rb$^{+}$  & 7 & 0.658 & 1.01 \\
 \end{tabular}
\end{ruledtabular}
\end{table}
%%%%%%%%%%%%%%%%%%%%%%%%%%%%%%%%%%%%%%%%%%%%%%%%%%%%%%%%%%%%%%%%%%%%%%%%%%%%%%%%%%%

%
%%%%%%%%%%%%%%%%%%%%%%%%%%%%%%%%%%%%%
\section{ Steric vs. dielectrophoretic Counter-ion Saturation}
\label{sec:saturations}
%%%%%%%%%%%%%%%%%%%%%%%%%%%%%%%%%%%%%%%%%%%%%%%%%%%%%%%%%%%%%%%%%%%%%%

At high surface-charge densities, the counter-ions accumulate at the
surface due to their strong electrostatic attraction to the oppositely charged surface. However,
at some point a saturation
concentration is reached as a result of two possible mechanisms: a
steric counter-ion saturation
or a dielectrophoretic one.
The steric effect is due to finite ion size and causes the
ionic concentration to saturate at closed packing,
estimated as $1/a^{3}$ where $a$ is the ionic size, while the dielectrophoretic
saturation is determined by the coupling between the local dielectric decrement and the ionic profile.

In the high surface-charge limit, we can
estimate the concentration at the dielectrophoretic
saturation using the following argument.  When the
dielectrophoretic saturation  is reached before closed packing of the ions,
the distance between ions is larger than the inter-ion closest approach and
the steric effect can be neglected in
Eq.~(\ref{eq:mpb_distribution}), such that $\rho_\pm=n_\pm$. In the
counter-ion plateau region, the condition for a counter-ion saturation given by $\partial n_{+}/\partial z=0$, leads also to $\partial \epsilon(n_{+}(z))/\partial z=0$.
Taking the $z$ derivative on both sides of Eq.~(\ref{eq:ppb_distribution}) yields
\begin{align}
\label{eq7}
0 &=\psi'\left(e -\varepsilon_{0}\frac{\partial\varepsilon}{\partial n_{+}}\psi''\right)\, .
\end{align}
For large enough $\left|\sigma\right|$, the co-ions are almost completely excluded from
the EDL, $n_{-}\approx 0$,
and the Poisson equation then becomes
\begin{align}
\label{eq8}
\varepsilon_{0} \varepsilon_s \psi''
&\approx -en_{s}^{+}\, ,
\end{align}
where $n_{s}^{\pm}=n_\pm(0)$ are the extrapolated ion concentrations at the surface, and
$\varepsilon_s=\varepsilon(n_{s}^{+},n_{s}^{-}=0)$.
Combining Eqs.~(\ref{eq7}) and (\ref{eq8}), we define a new function $\Delta\varepsilon_s(n_s^+)$
that should vanish at the saturation condition of
$n_{+}$,
\begin{align}
\Delta\varepsilon_s(n_s^+)\equiv  \varepsilon_s + n_{s}^{+}
\left.\frac{\partial\varepsilon}{\partial n_{+}}\right|_s = 0\, .
\label{eq:dielectrophoretic_saturation_condition}
\end{align}

In order to solve this equation and obtain $n_s^+$ at 
dielectrophoretic saturation, an explicit dependence of the permittivity on the
counter-ion concentration is required. The simplest model to employ
is the linear decrement model introduced
in Eq.~(\ref{eq:linear_decrement}), where
$\partial\varepsilon/\partial n_{\pm}=-\gamma_{\pm}$. As mentioned above, the linear
decrement model is a good approximation for bulk electrolytes of concentration
up to about 2\,M. Close to the
surface, where the counter-ion concentration can be quite high, one needs to
consider corrections to linearity as will be presented in
Sec.~\ref{sec:nonlinear_decrement} below.

Returning to the linear
decrement case, the above condition,
Eq.~(\ref{eq:dielectrophoretic_saturation_condition}), for 
dielectrophoretic saturation reduces to
\begin{align}
\label{eq10}
n_s^{+}&=\varepsilon_{w}/2\gamma_{+},\nonumber\\
\varepsilon_{s}&=\varepsilon_{w}/2 \, .
\end{align}
Note that although a linear decrement is assumed,
$\varepsilon_s$ at dielectrophoretic saturation does not approach zero for high
$n_{s}^{+}$.
The limiting values of Eq.~(\ref{eq10}) are identical to those derived earlier in the salt-free
case~\cite{BenYaakov2011Dielectric}, because even with added salt, the co-ions
are taken to be fully depleted from the highly charged surface proximity and do not contribute to the permittivity variation.

%%%%%%%%%%%%%%%%%%%%%%%%%%%%%%%%%%%%%%%%%%%%%%
\subsection{The generalized Grahame equation}
\label{sec:generalized_grahame}
%%%%%%%%%%%%%%%%%%%%%%%%%%%%%%%%%%%%%%%%%%%%%%%%%

More detailed analysis of counter-ion saturation at the surface can be obtained
through the contact theorem or Grahame
equation~\cite{2006Soft,Israelachvili2011Intermolecular}.
To derive the contact theorem, we first calculate the osmotic pressure
from the free energy, Eq.~(\ref{eq:free_energy})
\begin{align}
P(z) =&
-\frac{\varepsilon_{0}}{2} \left[
\varepsilon(n_\pm)+\sum_{\alpha=\pm}n_{\alpha}\frac{\partial
\varepsilon}{\partial n_{\alpha}}
\right] \left|\psi'\right|^{2}
\nonumber \\
& -\frac{k_{B}T}{a^{3}}\ln\phi_w(z) \, .
\label{P_os}
\end{align}
At equilibrium, $P(z)$ is a constant independent of $z$.
Equating the pressure at the charged surface $z=0$, with the pressure in
the bulk, $P(\infty)$, gives the modified Grahame equation.
Furthermore, when the co-ions are fully depleted from the
surface, $n_{s}^{-}\to 0$, and we obtain
\begin{align}
 \sigma^{2} &\simeq
\frac{\varepsilon_{0}\left(\varepsilon_{s}\right)^{2}}{\Delta\varepsilon_{s}}\frac{2k_{B}T}{a^{3}}
\ln \left(\frac{\phi_b}{\phi_s}\right) \, ,
\label{eq:pmpb_sigma_ns1}
\end{align}
where $\Delta\varepsilon_s$  was defined in
Eq.~(\ref{eq:dielectrophoretic_saturation_condition}),
$\phi_b=1-2a^{3}n_{b}$ and $\phi_s\simeq 1-a^{3}n_{s}^{+}$ are, respectively,
the solute (water) volume fraction
evaluated in the bulk and at the surface.
In the linear decrement case,
$\Delta\varepsilon_s$  is
\begin{equation}
\Delta\varepsilon_s=\varepsilon_s-\gamma_{+}n_s^{+}=\varepsilon_w-2\gamma_{+}n_s^{+}
\end{equation}
and the modified Grahame equation reduces to
\begin{align}
\label{mod_grahame}
\sigma^{2} &\simeq
\frac{\varepsilon_{0}\left(\varepsilon_{w}-\gamma_{+}n_{s}^{+}\right)^{2}}{
\varepsilon_{w}-2\gamma_{+}n_{s}^{+}
} \frac{2k_{B}T}{a^{3}}
\ln \left(\frac{\phi_b}{\phi_s}\right) \, .
\end{align}

When the permittivity does not depend at all on the ion concentrations, $\varepsilon(n_{\pm})=\varepsilon_w$,
the above Grahame equation recovers the sterically-modified Poisson--Boltzmann~(smPB)
result~\cite{Borukhov1997Steric}. It diverges logarithmically at ionic closed-packing, $n_{s}^{+}\to 1/a^{3}$,
which is the maximal value of the ion concentration.

From the generalized Grahame equation~(\ref{mod_grahame}),
it can be seen that the limiting saturation
value of $n_s^+$ is either
$\varepsilon_{w}/2\gamma_{+}$ for the dielectrophoretic saturation
or $1/a^{3}$ for closed packing.
As the surface charge density increases, $n_s^+$
reaches the smallest value between $\varepsilon_{w}/2\gamma_{+}$ and $1/a^{3}$.
To determine which of the two mechanisms prevails, we define an effective size
associated with the concentration at dielectrophoretic saturation,
\begin{equation}
\label{def_d}
d = \left(\frac{2\gamma}{\varepsilon_{w}}\right)^{1/3} \,.
\end{equation}
For ions with $d/a>1$ the dielectrophoretic saturation prevails,
while for ions with $d/a<1$ the sterically-dominant saturation predominates.
This ion-dependent parameter $d/a$ is shown in
Table~\ref{tbl:parameters} for several monovalent cations and anions.

\begin{figure}[htbp]
 %\center
\includegraphics{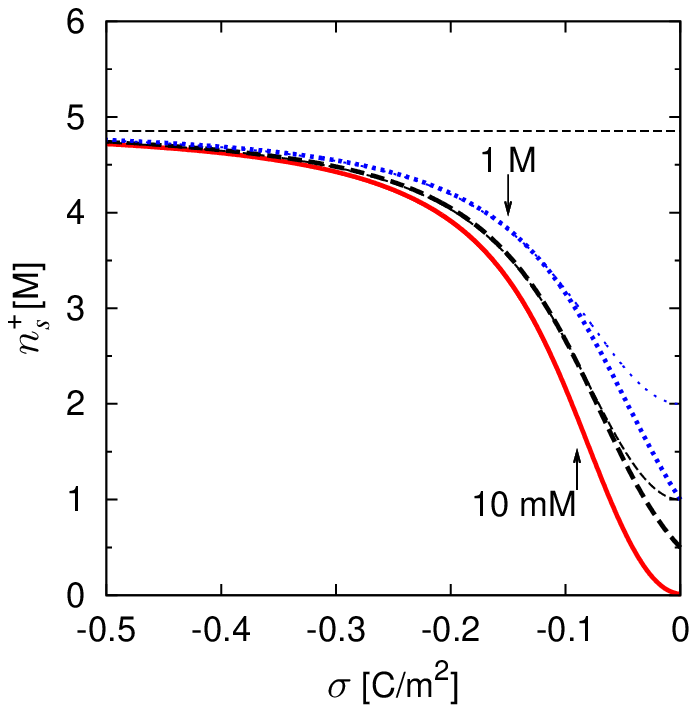}
\caption{\textsf{Dielectrophoretic saturation of counter-ion concentration dominated by the
K$^+$ counter-ions with $d/a\simeq 1.05$.
The counter-ion concentration, $n_{s}^+=n_{+}(0)$, at the proximity of a  charged surface, is plotted as
a function of the negative surface charge density, $\sigma<0$. The thicker lines are calculated numerically for
KCl with bulk values $n_b=$10\,mM~(solid red line), 0.5\,M~(dashed black line) and 1\,M (dotted blue line).
The  K$^+$ and
Cl$^{-}$ parameters are shown in Table~\ref{tbl:parameters},
$T=25^\circ$\,C and $\varepsilon_w=78.3$.
The thinner blue and black lines are the analytical approximated $n_{s}^+$ obtained from the Grahame
 equation~(\ref{mod_grahame}) with the same bulk values of $n_b$, respectively, as for the thick lines, and by setting $n_{s}^{-}=0$.
 Note that for the red line of 10\,mM, the analytical line is indistinguishable from the numerical one.
The concentration at the dielectrophoretic saturation,
 $\varepsilon_{w}/2\gamma_{\text{K}^{+}}\simeq 4.89$\,M, is indicated by a horizontal dashed line.
}}
\label{fig1}
\end{figure}

\begin{figure}[htbp]
 %\center
\includegraphics{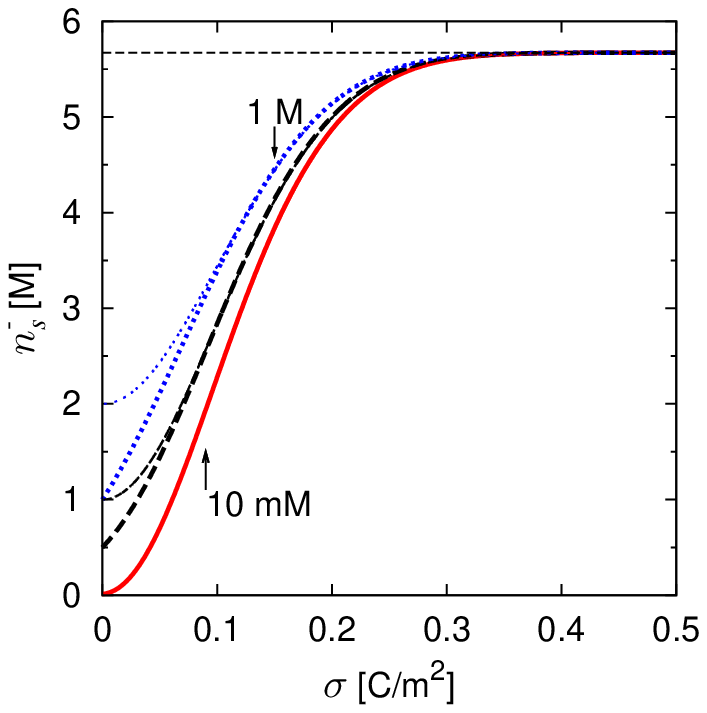}
\caption{\textsf{Sterical saturation of counter-ion concentration dominated by Cl$^-$ counter-ions with $d/a\simeq 0.76$.
The counter-ion concentration, $n_{s}^-=n_{-}(0)$, at the proximity of a charged surface, is plotted
as a function of the positive surface charge density, $\sigma>0$. The thicker lines are calculated numerically
for KCl with bulk values $n_b=$10\,mM~(solid red line), 0.5\,M~(dashed black line) and
 1\,M (dotted blue line).
The K$^+$ and
Cl$^{-}$ parameters are shown in Table~\ref{tbl:parameters},
$T=25^\circ$\,C and $\varepsilon_w=78.3$.
The thinner lines are the approximate $n_{s}^-$ as obtained from the Grahame
 equation~(\ref{mod_grahame}) with the same $n_b$ bulk values, respectively, and by setting $n_s^+=0$.
 Note that for the red line of 10\,mM, the analytical line is indistinguishable from the numerical one.
The concentration at closed packing,
 $1/a_{\text{Cl}^{-}}^{3}\simeq 5.67$\,M, is  indicated by a  horizontal dashed line.
}}
\label{fig2}
\end{figure}

As an example, we show in Fig.~\ref{fig1} results for potassium (K$^{+}$) ions
with the ratio $d/a\simeq 1.05>1$,
and in Fig.~\ref{fig2} results for chloride (Cl$^{-}$) ions  with the ratio
$d/a\simeq 0.76<1$ (ionic parameters are shown in
Table~\ref{tbl:parameters}).
The (extrapolated) surface value, $n_s^{\pm}=n_\pm(z\to 0)$, is plotted as function of $\sigma$ for several values of $n_b$.
For large $|\sigma|$,
the approach of $n_s^+$ towards the limiting value of
$\varepsilon_{w}/2\gamma_{+}\simeq 4.89$\,M is clearly observed~(Fig.~\ref{fig1} with $d/a>1$),
whereas  for $d/a<1$,  $n_{s}^{-}$ approaches $1/a^{3}\simeq 5.67$\,M as $\sigma$ increases
(Fig.~\ref{fig2}).
In Figs.~\ref{fig1} and \ref{fig2}, we also compare
Eq.~(\ref{mod_grahame}), where co-ions are completely depleted, with the full numerical calculation.
Indeed, a very good agreement is observed in the high $|\sigma|$ limit, confirming
the applicability of Eq.~(\ref{mod_grahame}) in this
limit, as expected.

%fig1
%%%%%%%%%%%%%%%%%%%%%%%%%%%%%%%%%%%%%%%%%%%%%%%%%%%%%%%%%%%%%%%%%%%
%
%%%%%%%%%%%%%%%%%%%%%%%%%%%%%%%%%%%%%%%%%%%%%%%%%%%%%%%%%%%%%%%%%%%

%fig2
%%%%%%%%%%%%%%%%%%%%%%%%%%%%%%%%%%%%%%%%%%%%%%%%%%%%%%%%%%%%%%%%%%%
%%%%%%%%%%%%%%%%%%%%%%%%%%%%%%%%%%%%%%%%%%%%%%%%%%%%%%%%%%%%%%%%%%%

%%%%%%%%%%%%%%%%%%%%%%%%%%%%%%%%%%%%%%%%%%%
\section{Differential capacitance}
\label{sec:difc}
%%%%%%%%%%%%%%%%%%%%%%%%%%%%%%%%%%%%%%%%%%%%%%
Irrespective of the underlying
mechanism, when counter-ion saturation  occurs, the EDL width will grow.  This effect can be shown by considering
the {\it differential capacitance}, $C(\psi_{s})=\partial
\sigma/\partial \psi_{s}$.

We derive an analytic expression for $C$ under
the assumption that co-ions are fully depleted from the surface, $n_s^{-}=0$.
As was shown earlier this gives a simplified
expression for the surface permittivity,
$\varepsilon_s \simeq\varepsilon_w - \gamma_{+}n_s^{+}$ and $\Delta \varepsilon_s \simeq\varepsilon_w - 2\gamma_{+}n_s^{+}$.
The results are presented here within the linear decrement model, and later (Sec.~\ref{sec:nonlinear_decrement})
they will be generalized to the non-linear case.
The relation between $\psi_s$ and $n_s^+$
is obtained from the equilibrium distribution,
Eqs.~(\ref{eq:mpb_distribution}) and (\ref{eq:ppb_distribution}), at the
surface by eliminating $\psi'$ with $\sigma(n_{s}^{+})$ from
Eq.~(\ref{mod_grahame}) and the boundary condition, Eq.~(\ref{eq:bc}), and $\rho_s^{+}=n_s^{+}(\phi_b/\phi_s)$
%

%%%%%%%%%%%%%%%%%%%%%%%%%%%%%%%%%%%%%%%%%%%%%%%%%%%%%%%
%
\begin{align}
\Psi_{s}
&=
\ln\left(\frac{n_{b}}{n_{s}^{+}}\right)
-\frac{\Delta\varepsilon_s+\gamma_{+}a^{-3}}
{\Delta\varepsilon_s}
\ln\left(\frac{\phi_{b}}{\phi_{s}}\right) \, .
\label{eq:psi_ns}
\end{align}
where $\Psi_s\equiv e\psi_s/k_BT$ is the dimensionless surface potential to be used hereafter.
Next, the differential capacitance is obtained as the parametric
function of $n_{s}^{+}$:

\begin{equation}
\label{C_parametric}
C= \frac{e}{k_B T}\frac{\partial \sigma}{\partial n_{s}^{+}}
\left(\frac{\partial \Psi_{s}}{\partial n_{s}^{+}}\right)^{-1},
\end{equation}
where
\begin{align}
\frac{\partial \Psi_{s}}{\partial n_{s}^{+}}
=&-\frac{1}{n_{s}^{+}}
-\frac{ \Delta\varepsilon_s+\gamma_{+}a^{-3}}{
\Delta\varepsilon_s} \left(\frac{a^{3}}{\phi_{s}}\right)
\nonumber\\
&\nonumber\\
&-\frac{2\gamma_{+}^2}{a^{3}
\left(\Delta\varepsilon_s\right)^{2}}
\ln\left(\frac{\phi_{b}}{\phi_{s}}\right)\, ,
\label{eq:dpsi_dns}
\end{align}
and

\begin{align}
\frac{\partial\sigma}{\partial n_{s}^{+}}
=&\frac{1}{\Delta\varepsilon_s}
\left(\frac{k_{B}T\varepsilon_{0}\varepsilon_{s}^2}{\phi_s\sigma}
 +\frac{\sigma n_{s}^{+}\gamma_{+}^2}{\varepsilon_{s}} \right)\, ,
\label{eq:dsigma_dns}
\end{align}
%

%fig3
%%%%%%%%%%%%%%%%%%%%%%%%%%%%%%%%%%%%%%%%%%%%%%%%%%%%%%%%%%%%%%%%%%%%%%%
\begin{figure}[htbp]
 %\center
\includegraphics{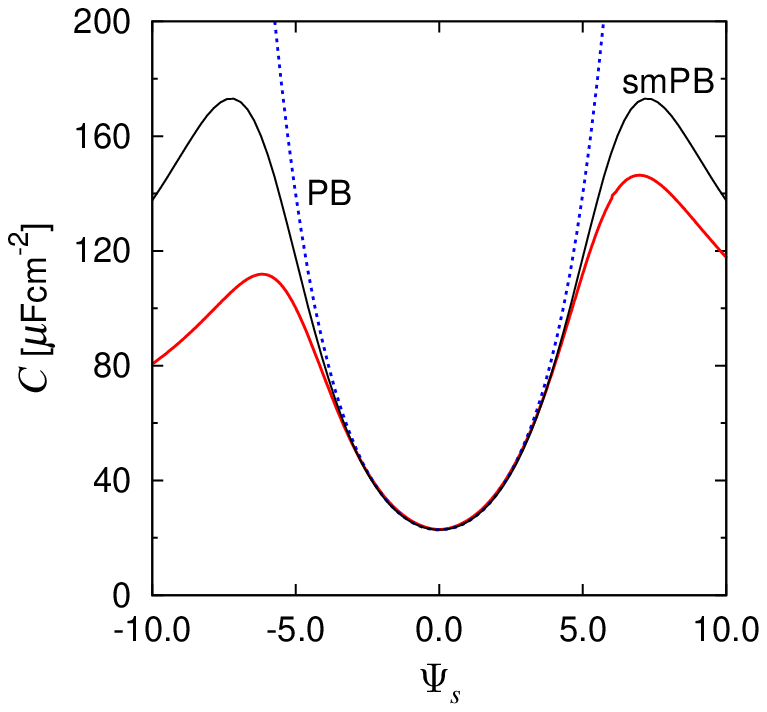}
\caption{\textsf{
Comparison of the differential capacitance $C$ calculated for
KCl (parameters as in Table~\ref{tbl:parameters}) with $n_{b}=10$\,mM.  The red solid
line corresponds to the exact numerical solution, the blue dotted line is
the regular PB result, Eq.~(\ref{C_PB}), and the black solid line is the smPB result,
Eq.~(\ref{eq:difc_smpb}).
The peaks of the red curves are located at
$\Psi_s=6.98$ and
$\Psi_s=-6.15$, and those of the black curves are at
$\Psi_s=\pm 7.25$.
}}
\label{fig3}
\end{figure}
%%%%%%%%%%%%%%%%%%%%%%%%%%%%%%%%%%%%%%%%%%%%%%%%%%%%%%%%%%%%%%%%%%%%%%%%%%%%

%fig4
%%%%%%%%%%%%%%%%%%%%%%%%%%%%%%%%%%%%%%%%%%%%%%%%%%%%%%%%%%%%%%%%%%%%%%%%%%%%
\begin{figure}[htbp]
 %\center
\includegraphics{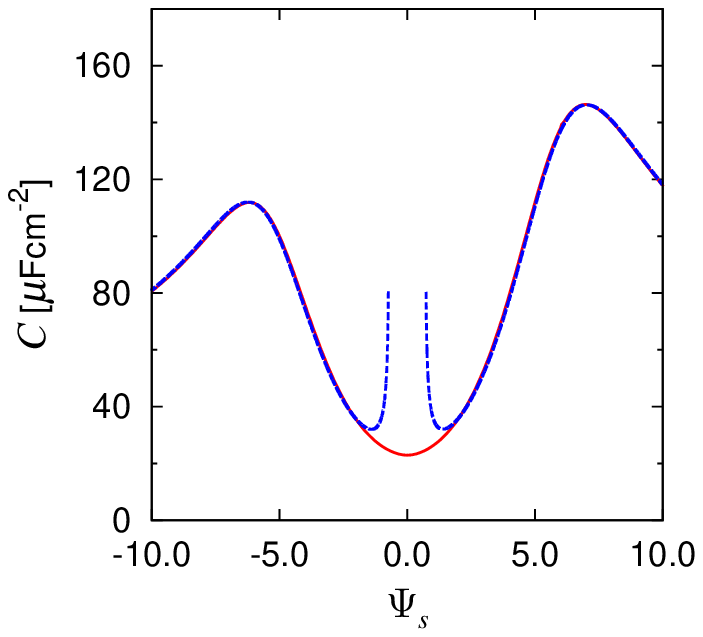}
\\
\caption{\textsf{
Comparison of the analytic differential capacitance $C$ with the numerical one calculated for
KCl (parameters as in Table~\ref{tbl:parameters}) with $n_{b}=10$\,mM.  The red solid
line (just as in Fig.~\ref{fig3}) corresponds to the full numerical solution, while the blue dashed line
denotes the approximate analytic expression of
Eqs.~(\ref{eq:psi_ns})-(\ref{eq:dsigma_dns}). The analytical results reproduce the numerical ones in all range
of
$|\Psi_s|\gtrsim 2$, including the two asymmetric peak heights at
$\Psi_s=6.98$ and
$\Psi_s=-6.15$, while $C_{\rm PB}$ reproduces well the $C$ behavior for
 small $|\Psi_s|\lesssim 2$.
}}
\label{fig4}
\end{figure}

Figures~\ref{fig3} and \ref{fig4} shows the numerically-calculated
$C$ as a function of $\Psi_{s}$ for KCl concentration $n_{b}=10$\,mM, along with
several analytic results.  At this salt concentration, the calculated
$C$ shows two peaks, one for $\Psi_s< 0$ and the other for $\Psi_s>0$.  This
characteristic shape of the differential capacitance is called {\it
camel-shape} or {\it double-hump}~\cite{Kornyshev2007DoubleLayer},
and is often observed in experimental differential
capacitance~\cite{Grahame1954Differential,Valette1981Double,Valette1982Double,Clavilier1977Etude}
at relatively low salt concentrations.
It is quite different from
the well-known capacitance of the standard PB model, $C_{\rm PB}$, which is
obtained in our  model by setting
$\gamma_{\pm}=a=0$,
\begin{equation}
C_{\text{PB}}=\varepsilon_{0}\varepsilon_{w}\lambda_D^{-1}\cosh(\Psi_{s}/2) \, ,
\label{C_PB}
\end{equation}
where
$\lambda_{D} = \sqrt{\varepsilon_{0}\varepsilon_{w}k_{B}T/2n_{b}e^{2}}$
is the Debye-H\"uckel screening length.  The PB capacitance is
applicable to small surface potentials, and has a minimum at $\Psi_s=0$. However, as is seen in Fig.~\ref{fig3},
it does not reproduce the two peaks for larger $\left|\Psi_s\right|$.
In contrast, the capacitance calculated from the analytical expressions,
Eqs.~(\ref{eq:psi_ns})-(\ref{eq:dsigma_dns}), is found to reproduce well
the full (numerical) dependence of $C(\Psi_{s})$, including the two
asymmetric peaks in $C$~(Fig.~\ref{fig4}).  A deviation of this analytical expression from the numerical one
is noticed at small values of the potential,
$|\Psi_s|\lesssim 2$, and
$C$ from Eqs.~(\ref{eq:psi_ns})-(\ref{eq:dsigma_dns}) even diverges for $|\Psi_{s}|\to 0$,
because the assumption that co-ions are fully depleted from the
surface is not valid anymore.

It is possible to describe the full dependence of $C=C(\Psi_s)$
by two  analytical expressions: { (i)}~$C_{\text{PB}}$ of
Eq.~(\ref{C_PB}) for small $\Psi_{s}$,
and { (ii)}~the expression of
Eqs.~(\ref{eq:psi_ns})-(\ref{eq:dsigma_dns})  for large
$\Psi_{s}$. Using the approximate two-patch expressions for $C$, we can reproduce
a line that is almost indistinguishable from the red line  of Fig.~\ref{fig4},
which was obtained numerically from the full expression.
The crossover between the two regions occurs at about the same $\Psi_s$ where $C$
(derived from Eqs.~(\ref{eq:psi_ns})-(\ref{eq:dsigma_dns}))
starts to sharply increase above $C_{\rm PB}$.
Below the crossover $\Psi_s$, the co-ion concentration at the surface,
$n_{s}^{-}$,  becomes significant. As $n_{s}^{-}$
reduces the capacitance, Eqs.~(\ref{eq:psi_ns})-(\ref{eq:dsigma_dns})
that {\it do not} include their presence, start to deviate sharply from the behavior
describes by $C_{\rm PB}$.

We note that the two peaks at the two surface-polarities, $\Psi_{s}\lessgtr 0$,
have different origins resulting in non-equal peak heights.
The build-up of the EDL
is dominated by the cations for $\Psi_{s}<0$ and by
the anions for $\Psi_{s}>0$. In the particular example
for KCl (Fig.~\ref{fig3}),  K$^+$ is
the counter-ion for $\Psi_{s}<0$  with $d/a>1$ (dielectrophoretic-dominant case), while Cl$^-$ is
the counter-ion for $\Psi_{s}>0$ with $d/a<1$ (sterically-dominant case), and this explains the asymmetry  of the two peaks.

In order to check the effect of dielectric decrement,
we also plot in Fig.~\ref{fig3} the analytic differential capacitance for the smPB
model, $C_{\rm smPB}$, which was  derived previously~\cite{Kornyshev2007DoubleLayer,Kilic2007Steric_i}
\begin{align}
C_{\text{smPB}} =&
\frac{C_{\text{PB}}}
{1+4a^{3}n_{b}\sinh^{2}(\Psi_{s}/2)}\nonumber\\
&\nonumber\\
& \times
\sqrt{\frac{4a^{3}n_{b}\sinh^{2}(\Psi_{s}/2)}
{\ln\left(1+4a^{3}n_{b}\sinh^{2}(\Psi_{s}/2)
\right)}} \, ,
\label{eq:difc_smpb}
\end{align}
The peak position within the smPB model can  be roughly estimated~\cite{Kilic2007Steric_i,Kilic2007Steric_ii} by
substituting the closed-packing concentration, $n_{s}=1/a^{3}$ into the
Boltzmann distribution,
$n_{s}^{\pm}=n_{b}\exp\left(\mp\Psi^{\text{str}}\right)$, yielding
$\left|\Psi^{\rm str}\right| \simeq
-\ln\left(a^{3}
n_{b}\right)$.
For the case of Fig.~\ref{fig3}, it gives
$\left|\Psi^{\text{str}}\right|\simeq6.34$ with $a_{\text{Cl}^{-}}=0.664$\,nm.
Although $C_{\text{smPB}}$ also exhibits double-hump shape,
the peak positions (in $\Psi_{s}$) and their corresponding heights differ
from the ones in our
model, as will be discussed in detail in the subsequent sections.

For KCl with \(n_{b}=\)10\,mM as in Fig.~\ref{fig3}, \(C\) at \(\Psi_{s}=0\) is
22.9\,\(\mu\)Fcm\(^{-2}\), and it overestimates typical experimental
data~\cite{Bonthuis2013Beyond}.
This discrepancy is probably due to the omission  of the Stern layer.
For quantitative description of experimental differential capacitance,
appropriate modeling of the Stern layer is
required in addition to the EDL.

\begin{figure}[htbp]
 %\center
\includegraphics{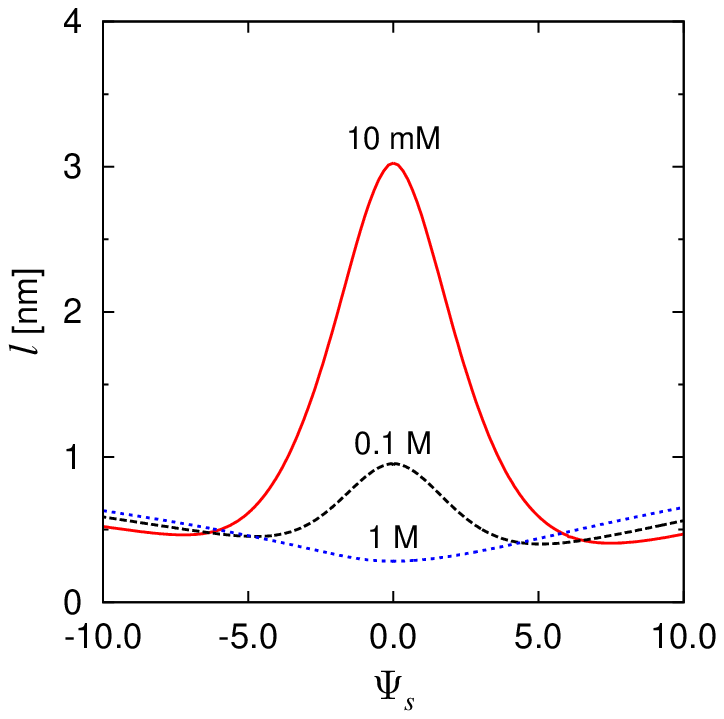}
\caption{\textsf{The width of the
electric double layer estimated from the differential capacitance,
$l=\varepsilon_{0}\varepsilon_{s}/C$, as a function of
dimensionless surface potential, $\Psi_{s}$, for KCl (parameters as in
Table~\ref{tbl:parameters}) with $n_b=$10\,mM~(solid red line),
0.1\,M~(dashed black line) and 1\,M (dotted blue line).  }}
\label{fig5}
\end{figure}

From the differential capacitance, we can estimate the EDL width as
\begin{equation}
l(n_{b},\sigma(\psi_{s}))= \varepsilon_{0}\varepsilon_{s}/C \, .
\end{equation}
In
Fig.~\ref{fig5}, $l$ is plotted as a function of $\Psi_{s}$
for several values of $n_{b}$.
For small $n_{b}$ ($n_{b}=10$\,mM in Fig.~\ref{fig5}), $l$ first decreases with
$|\Psi_{s}|$ due to the increasing electrostatic attraction between the surface
and the counter-ions. Note that the regular PB
theory is valid for such small $\left|\Psi_{s}\right|$, and gives similar results.
For larger $\left|\Psi_{s}\right|$, after saturation of the counter-ion concentration is
reached,
$l$ starts to increase with $|\Psi_{s}|$
as $n_{s}$ reached saturation and the accumulated counter-ions
contribute to thickening the EDL.

As $n_{b}$ increases ($n_{b}=0.1$\,M in Fig.~\ref{fig5}), $n_{s}$
can reach  saturation at even
smaller $|\Psi_{s}|$, and the range of $\Psi_{s}$ where
the PB theory can be applied diminishes.
When $n_{b}$ increases even further ($n_{b}=1$\,M in Fig.~\ref{fig5}),
$n_{s}$ immediately reaches saturation
even for small applied $\Psi_{s}$. Therefore, the increase of $l$ with
$\left|\Psi_{s}\right|$ already starts from $\Psi_{s}=0$ .  In this high salinity range, the regular PB theory
cannot be applied at all.  The crossover from low- to high-salinity will be
further discussed in Sec.~\ref{sec:high_salinity}.

%fig5
%%%%%%%%%%%%%%%%%%%%%%%%%%%%%%%%%%%%%%%%%%%%%%%%%%%%%%%%%%%%%%%%%%%
%%%%%%%%%%%%%%%%%%%%%%%%%%%%%%%%%%%%%%%%%%%%%%%%%%%%%%%%%%%%%%%%%%%%%%%%%%%%

%%%%%%%%%%%%%%%%%%%%%%%%%%%%%%%%%%%%%%%%%%%%%%%%%%%%%%
\subsection{The dielectrophoretic saturation for $d>a$ }
\label{sec:dielectrophoretic_saturation}
%%%%%%%%%%%%%%%%%%%%%%%%%%%%%%%%%%%%%%%%%%%%%%%%%%%%%%%

For the case of KCl in contact with a negatively charged surface, the
counter-ion is K$^{+}$ (potassium) with $d/a>1$, where \(d\)
has been already defined in Eq.~(\ref{def_d}) as the effective size at the
dielectrophoretic saturation concentration. The EDL of this
counter-ion exhibits a dielectrophoretic saturation.  The location of the peak in $C$, $\Psi^{\rm
die}<0$, can be estimated similarly to the way $\Psi^{\rm str}$ was
estimated after Eq.~(\ref{eq:difc_smpb}) above.  We substitute $n_s^{+}=
\varepsilon_{w}/2\gamma_{+}$ into the Boltzmann distribution, and obtain
$\Psi^{\rm die}=e\psi^{\rm die}/k_BT \simeq
-\ln(\varepsilon_{w}/2\gamma_{+} n_{b})$.  For the case in
Fig.~\ref{fig3}, this estimation gives $\Psi^{\text{die}}\simeq -6.19$.

\begin{figure}[htbp]
\center
\includegraphics{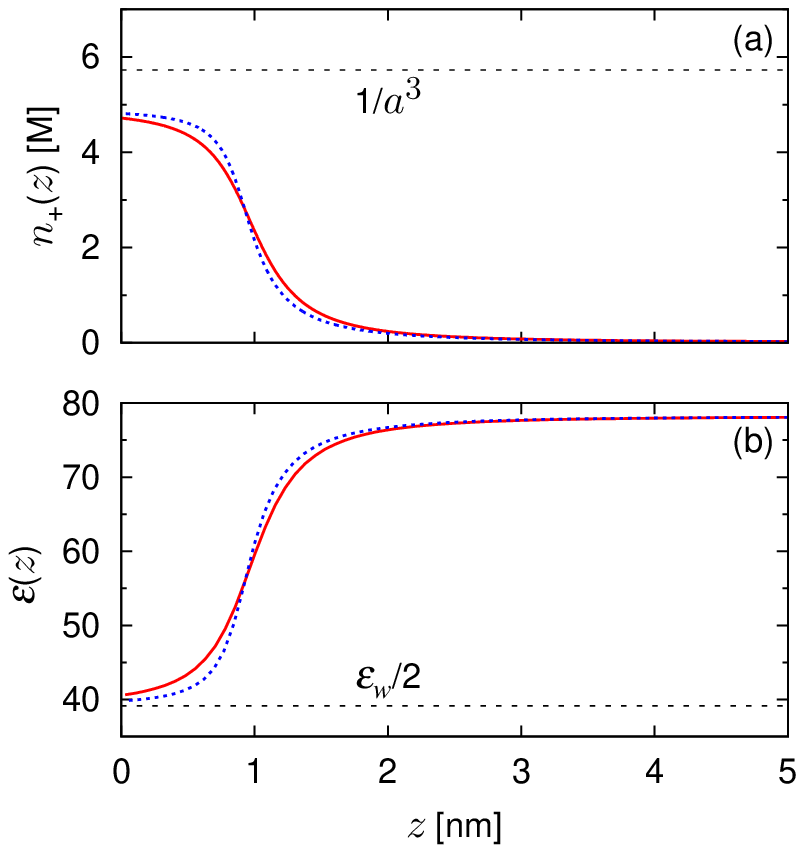}
\caption{\textsf{
(a) Counter-ion density profile, $n_{+}(z)$,  as function of the distance $z$ from a
negatively charged surface with
$\sigma=-0.5$\,C/m$^{2}$, for KCl concentration $n_{b}=10$\,mM~(red solid
line). The K$^+$ values of $a$ and $d$ are taken from Table~I. The dotted (blue) line represents the profile of the regular PB
that includes the dielectrophoretic effect, $\varepsilon=\varepsilon(n(z))$, but without the steric effect ($a=0$).  The dashed (black) line
corresponds to  closed-packing concentration $1/a^3=5.72$\,M
for K$^{+}$, and is bigger than the actual saturation concentration here.  (b) The
relative permittivity profile, $\varepsilon(z)$~(red solid line). The dotted blue line represents the profile of the regular PB
that includes the dielectrophoretic effect.  The
dashed line corresponds to the permittivity at the limiting
dielectrophoretic saturation, $\varepsilon_s=\varepsilon_w/2$.
}}
\label{fig6}
\end{figure}

For ions with $d/a>1$, the threshold surface potential for
dielectrophoretic counter-ion saturation is
smaller than that for steric saturation. Namely, $\left|\Psi^{\rm
die}\right|<\left|\Psi^{\rm str}\right|$, and the peak in $C$ originates
from the dielectrophoretic counter-ion saturation.
In Fig.~\ref{fig6}, we plot the K$^{+}$ counter-ion profile,
$n_+(z)$, and the corresponding permittivity variation,
$\varepsilon(z)$, as function of the distance $z$ from the surface, for $n_{b}=10$\,mM.
For this case~\cite{BenYaakov2011Dielectric,Hatlo2012Electric} with $d/a>1$,  $n_s^+$
approaches the concentration at the dielectrophoretic saturation, which is smaller
than the closed-packing concentration, $\varepsilon_{w}/2\gamma_{+} <1/a^3$ (see Fig.~\ref{fig6}(a)).

For comparison, the results of the regular PB model ($a=0$), which, nevertheless, includes the
dielectrophoretic effect
via $\varepsilon(n(z))$, are also plotted in Fig.~\ref{fig6} (dotted blue line).
The steric effect slightly suppresses the concentration near the
surface, but the width of the diffuse layer is almost unaffected.  The
surface permittivity $\varepsilon_s$ approaches $\varepsilon_{w}/2$,
which is half that of the bulk. Hence, the
differential capacitance for counter-ions with $d/a>1$ is considerably
lower than that derived for the smPB model. In Fig.~\ref{fig3}, the
$C$-value at the $\Psi_{s}<0$ peak of the full model is 111.9
$\mu$Fcm$^{-2}$, while that of the smPB is 173.1 $\mu$Fcm$^{-2}$.

For $\left|\Psi_{s}\right|>\left|\Psi^{\rm die}\right|$, the differential
capacitance starts to decrease due to the increasing EDL width, $l$.
At very high surface potentials, $\left|\Psi_{s}\right|\gg \left|\Psi^{\rm die}\right|$, the
asymptotic form of the differential capacitance for the
dielectrophoretic saturation can be derived from Eqs.~(\ref{eq:dpsi_dns})-(\ref{eq:dsigma_dns}):
\begin{align}
C^{\infty}_{\rm die} &\simeq \sqrt{
\frac{\varepsilon_{0}\varepsilon_{w}e^{2}}{4d^{3}k_{B}T}}
\frac{1}{\sqrt{\left|\Psi_{s}\right|+\ln \left(d^{3}n_{b}\right)}} \, ,
\end{align}
The inverse-square-root decay of $C$ is similar to that observed in the
sterically-dominant saturation~\cite{Kornyshev2007DoubleLayer,Kilic2007Steric_i}, but
with the difference that the effective size $d$ replaces $a$, and the
limiting surface permittivity $\varepsilon_{s}= \varepsilon_{w}/2$
replaces $\varepsilon_{w}$.

As a conclusion of the above discussion
we would like to state that for ions with $d/a>1$, the
stronger mechanism is the dielectric decrement, and
the peak positions of the differential capacitance
occur at lower values of $\left|\Psi_{s}\right|$ than for the smPB model.
The
peak heights in $C$ are reduced because of the dielectrophoretic
saturation and corresponding reduction of the surface permittivity.

%
%fig6
%%%%%%%%%%%%%%%%%%%%%%%%%%%%%%%%%%%%%%%%%%%%%%%%%%%%%%%%%%%%%
%%%%%%%%%%%%%%%%%%%%%%%%%%%%%%%%%%%%%%%%%%%%%%%%%%%%%%%%%%%%%%%%%%%%%%%%%%%%%%%%%%%

%%%%%%%%%%%%%%%%%%%%%%%%%%%%%%%%%%%%%%%%%%%%%%%%%%%%%%%%%%%%%%%%
\subsection{The sterically-dominant saturation for $d<a$}
\label{sec:steric_saturation}
%%%%%%%%%%%%%%%%%%%%%%%%%%%%%%%%%%%%%%%%%%%%%%%%%%%%%%%%%%%%%%%

For ions with $d/a<1$, the hydration size, $a$, is
larger than the effective size at dielectrophoretic saturation,
$d$, the magnitude of the threshold surface potential
is smaller for the sterically-dominant saturation of
counter-ion concentration than for the dielectrophoretic one. Namely,
$\left|\Psi^{\rm str}\right|<\left|\Psi^{\rm die}\right|$, and the peak
in $C$ originates from the sterically-dominant counter-ion saturation.  This
is seen for large values of $\Psi_s>0$ in Fig.~\ref{fig3}, where the
behavior in $C$ is dominated by the Cl$^{-}$ counter-ion with $d/a<1$,
and the EDL exhibits the sterically-dominant counter-ion saturation (Fig.~\ref{fig2}).

\begin{figure}[htbp]
 \center
\includegraphics{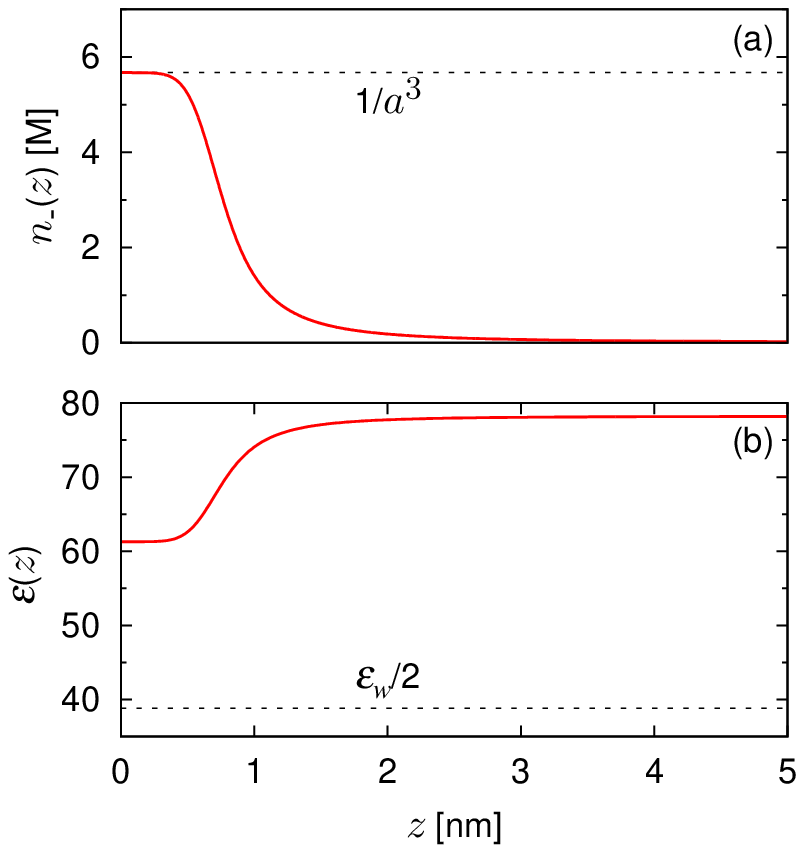}
\caption{\textsf{
(a) Counter-ion density profile, $n_{-}(z)$, as function of the distance $z$ from a positively charged surface, $\sigma=0.5$\,C/m$^{2}$,
for KCl concentration $n_b=10$\,mM~(red solid line). The Cl$^{-}$ values of $a$ and $d$ are taken from Table~I.
The dashed line corresponds to closed-packing concentration
for Cl$^{-}$, $n_s^{-}=1/a^3\simeq 5.67$\,M.
(b) Corresponding permittivity profile, $\varepsilon(z)$~(red solid line).
The dashed line corresponds to the permittivity at the limiting
 dielectrophoretic saturation, $\varepsilon_w/2$.
}}
\label{fig7}
\end{figure}

In Fig.~\ref{fig7}, we plot the Cl$^{-}$
counter-ion profile, $n_{-}(z)$, and the corresponding permittivity
variation, $\varepsilon(z)$, for $n_{b}=10$\,mM.
At the surface, $n_s^{-}$ reaches  the closed-packing
value of $ 1/a^{3}$, because the latter is smaller than the
dielectrophoretic value, $1/a^{3}<\varepsilon_{w}/2\gamma_{-}$.  For
comparison, the results of the smPB model with $\gamma_\pm=0$ are
also calculated.
However, it turns out that the ion density profiles with or without
$\gamma_{\pm}$ are almost identical (and their slight difference is below the resolution of the figure).
As the Cl$^{-}$ concentration is below the concentration at the
dielectrophoretic saturation for $d/a<1$, the dielectric
decrement is smaller~(Fig.~\ref{fig7}(b)), as compared with the K$^+$ case of Fig.~\ref{fig6}(b).

Although the counter-ion concentration, $n_{-}(z)$, is barely affected by
the dielectric decrement, the smPB model overestimates the peak position (in $\Psi_s$)
and the height of $C$~(Fig.~\ref{fig3}). The dielectric decrement in the diffuse
layer (with $\varepsilon$ decreasing from $78.3$ to about $61.3$) is
not so large, but the effect of dielectric decrement prevails even for $d/a<1$.
For the sterically-dominant saturation, the asymptotic expression for
$\left|\Psi_s\right|\gg \left|\Psi^{\rm str}\right|$
can be derived from
Eqs.~(\ref{eq:dpsi_dns})-(\ref{eq:dsigma_dns}):
\begin{align}
 C^{\infty}_{\rm str} &\simeq\sqrt{
\frac{e^{2}}{2a^{3}k_{B}T}}
\sqrt{\frac{\varepsilon_{0}\left(\varepsilon_{w}-\gamma_{-}/a^{3}\right)}
{\left|\Psi_{s}\right|+\ln (a^{3}n_{b})}}\, .
\label{eq:difc_steric_asymptotic}
\end{align}
%%%%%%%%%%%%%%%%%%%
In comparison~\cite{Kornyshev2007DoubleLayer,Kilic2007Steric_i},
the asymptotic expression of $C_{\text{smPB}}$ from
Eq.~(\ref{eq:difc_smpb}) is just like Eq.~(\ref{eq:difc_steric_asymptotic})
but with $\gamma_{-}=0$.
As compared to the smPB model, $C^{\infty}_{\rm str}$ is reduced by the effect
of the permittivity decrement at the surface, $\varepsilon_s<\varepsilon_w$.

%fig7
%%%%%%%%%%%%%%%%%%%%%%%%%%%%%%%%%%%%%%%%%%%%%%%%%%%%%%%%%%%%%%%%%%%%

%%%%%%%%%%%%%%%%%%%%%%%%%%%%%%%%%%%%%%%%%%%%%%%%%%%%%%%%%%%%%
\subsection{Crossover from camel-shape to bell-shape at high salt}
\label{sec:high_salinity}
%%%%%%%%%%%%%%%%%%%%%%%%%%%%%%%%%%%%%%%%%%%%%%%%%%%%%%%%%%%%%
As the salt concentration $n_{b}$ increases, the threshold in $|\Psi_s|$
either for the dielectrophoretic saturation, $\left|\Psi^{\rm
die}\right|$, or for the sterically-dominant saturation,
$\left|\Psi^{\rm str}\right|$, decreases and the position of the two
peaks in $C$ approaches $\Psi_s=0$.  Simultaneously, the differential
capacitance at $\Psi_{s}=0$, estimated as $C(0)\approx
C_{\text{PB}}(0)=\varepsilon_{0}\varepsilon_{w}/\lambda_{D}$,
increases as $n_{b}$ increases, until at some value of $n_{b}$,
$C$ becomes uni-modal (bell shape) with a single peak at $\Psi_s\approx 0$.
Clearly that for these high salinity conditions, the standard
PB is not valid even for small $\Psi_s$.

We would like to get an estimation for $n_{b}$ at the crossover from
camel-shape to bell-shape.
At this $n_{b}^*$ value, the differential capacitance at $\Psi_{s}=0$
becomes comparable to the lower of the two peaks (located at $\Psi_s\lessgtr 0$).
The $C$-values at the two peaks can be roughly estimated by using the expression of $C_{\rm PB}$
of Eq.~(\ref{C_PB}), where $\varepsilon_{w}$ in $C_{\text{PB}}$ is replaced with
$\varepsilon_{s}$:
\begin{eqnarray}
 %\,.\label{c_cross}
C(\Psi_s)&=&
\sqrt{\frac{2\varepsilon_0\varepsilon_s e^2n_b}{k_B T}}\cosh(\Psi_s/2)\,.\label{c_cross}
\end{eqnarray}

For ions with $d/a>1$ where the dielectrophoretic saturation is
dominant, $C$ at the peak is estimated by substituting
$|\Psi^{\text{die}}|\simeq \ln(\varepsilon_w/2\gamma n_b)$ and
$\varepsilon_s=\varepsilon_w/2$ into Eq.~(\ref{c_cross})
\begin{align}
 C(\Psi^{\text{die}}) &\simeq \frac{C(0)}{\sqrt{8}}\sqrt{\frac{\varepsilon_{w}}{2\gamma n_{b}}} \, .
\end{align}
Namely, for ions with
$d/a>1$ the crossover is expected to occur around
$n_{b}^*\simeq \varepsilon_{w}/16\gamma$, and for K$^{+}$ it gives
$n_{b}^*\simeq 0.61$\,M.

For the other case of ions with $d/a<1$, where the steric saturation is
dominant, $C$ at the peak is estimated by substituting
$\Psi^{\text{str}}\simeq -\ln(a^3 n_b)$ and
$\varepsilon_{s}=\varepsilon_{w}-\gamma/a^{3}$  into the $C$ expression
from Eq.~(\ref{c_cross})
\begin{align}
 C(\Psi^{\text{str}}) &\simeq C(0)
\sqrt{
\frac{2-(d/a)^{3}}{8a^{3}n_{b}}
}\, .
\end{align}
and the crossover salt concentration occurs at:
\begin{align}
n_{b}^* &\simeq
\frac{2-({d}/{a})^{3}}{8 a^{3}} \, .\label{n_b*}
\end{align}
For Cl$^{-}$,
this salt concentration is estimated to be $n_{b}^*\simeq 1.1$\,M.
Therefore, combining the results for K$^+$ and Cl$^-$, we find that
the crossover from camel-shape to
bell-shape for KCl is expected at the smallest of the $n_{b}^*$  estimates for K$^+$ and Cl$^-$. And in our case,
$n_{b}^{*}\lesssim 0.61 $\,M.

\begin{figure}[htbp]
\center
\includegraphics{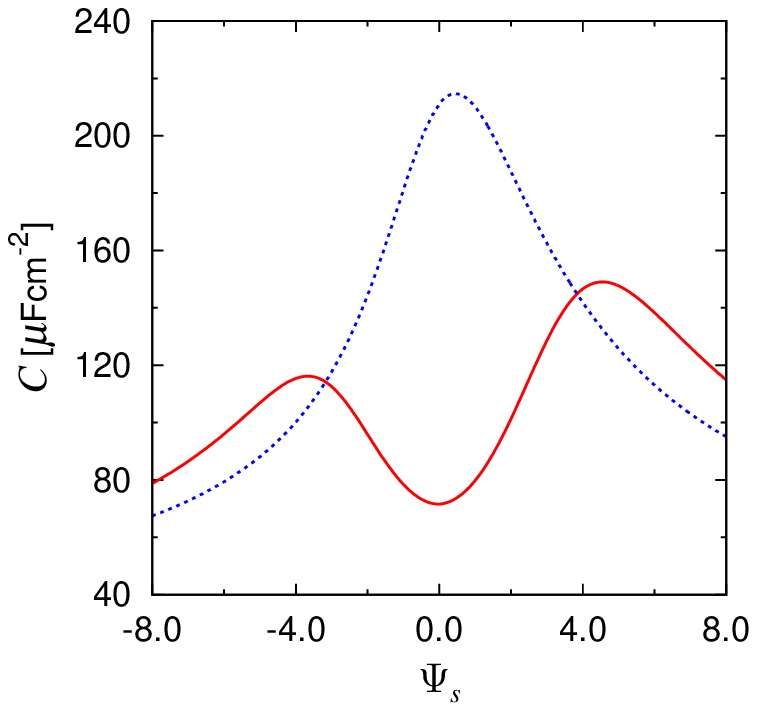}
\\
\caption{ \textsf{
Differential capacitance $C$ as a function of dimensionless surface potential, $\Psi_{s}$,
calculated for KCl (parameters in Table~\ref{tbl:parameters}).
For $n_{b}=0.1$\,M (red solid line)
$C$ exhibits a double-humped camel-shape, while for $n_{b}=1$\,M (blue dotted
line) it exhibits skewed bell-shape. The peaks of the red line are
 located at
$\Psi_s=4.56$
and
$\Psi_s=-3.67$, while the peak of the uni-modal $C$ occurs at
$\Psi_s=0.46$.
}}
\label{fig8}
\end{figure}

In Fig.~\ref{fig8}, the numerically calculated
$C$ for KCl concentration of $n_{b}=0.1$\,M and 1\,M
is plotted.
For the lower salt concentration of 0.1\,M, a double-hump $C$ is
observed, while for the higher salt concentration of 1\,M,
$C$ becomes uni-modal, as the two peaks merge together.
We remark that for the smPB model, the uni-modal (bell-shape) differential capacitance has been already derived
in Ref.~\onlinecite{Kornyshev2007DoubleLayer}, while in
the present work the changeover from double-hump to
bell-shape capacitance is investigated, by considering the combined ion finite-size and
dielectric decrement effects.
The value of $n_{b}^*$ at the crossover, as obtained from our numerical
calculations, is about 0.6\,M, which is comparable to $n_{b}^{*}$ estimated above.

For high salinity, the bell-shaped $C$ is skewed and its peak is located at slightly positive
surface potential, $\Psi_s\gtrsim 0$ (Fig.~\ref{fig8} with $n_b=1$\,M),
in contrast to lower salt concentration, where the
minimum in $C$ always occurs at $\Psi_{s}=0$.
When the bell-shaped $C$ occurs at high $n_{b}$, the
EDL width increases with $\left|\Psi_{s}\right|$. Consequently,
it always contributes to a decrease in $C$ as
$\left|\Psi_{s}\right|$ increases, because of the relation, $C=
\varepsilon_{0}\varepsilon_{s}/l$.
Accordingly, the increase of $C$
implies an increase of $\varepsilon_s$ with
$\left|\Psi_{s}\right|$.

\begin{figure}
\includegraphics{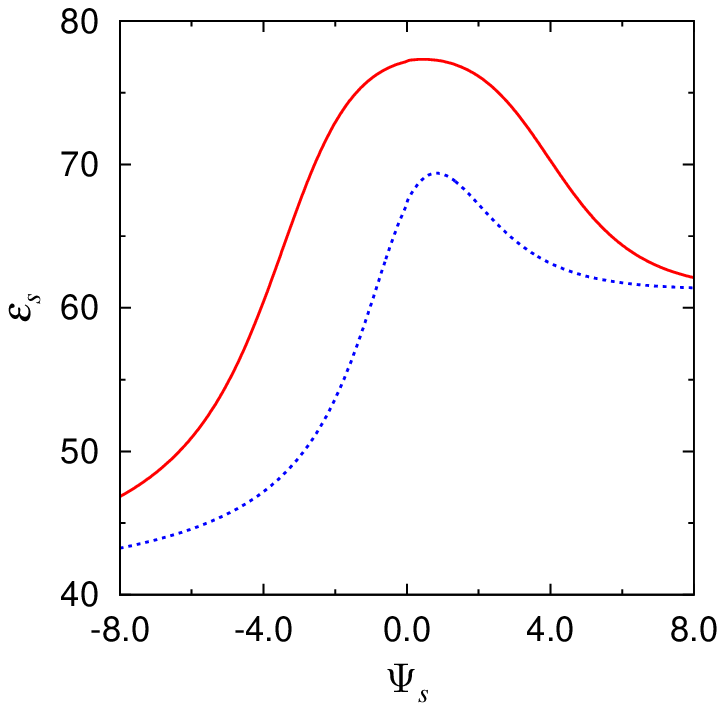}
\caption{\textsf{
Surface permittivity, $\varepsilon_{s}$, as
a function of the dimensionless surface potential, $\Psi_{s}$, for KCl
 (parameters in Table~\ref{tbl:parameters}).
The solid (red) line corresponds to $n_{b}=0.1$\,M, with a peak at
$\Psi_s=0.46$, while the dotted (blue) line corresponds to $n_{b}=1$\,M, with a peak at
$\Psi_s=0.81$.
}}
\label{fig9}
\end{figure}

In Fig.~\ref{fig9}, $\varepsilon_{s}$ is
plotted as a function of $\Psi_{s}$. The figure clearly shows that
for $\Psi_{s}>0$, $\varepsilon_{s}$ first increases from
$\Psi_{s}=0$, then has a peak at small and finite $\Psi_{s}$, and afterwards it decreases. From
this observation, we assert that the peak of the bell-shaped $C$
originates from the increase of $\varepsilon_{s}$ for small $\Psi_{s}>0$.
When finite $\left|\Psi_{s}\right|$ is applied, an EDL develops and
the counter-ion $n_{s}$ increases. However, it does not always
mean that $\varepsilon_{s}$ decreases with $\left|\Psi_{s}\right|$.
With the built-up of EDL, $n_{s}$ of the co-ions decreases, which
contribute to {\it increasing} $\varepsilon_{s}$ from the value
$\epsilon_{w}-(\gamma_{+}+\gamma_{-})n_{b}$ at $\Psi_{s}=0$.
To understand the observed non-monotonic change of $\varepsilon_{s}$ with
$\Psi_{s}>0$, the asymmetry of $\gamma_{\pm}$ should be considered,
in addition to the ionic concentration change between the bulk and the surface, $n_{s}^{\pm}-n_{b}$.

For the specific case of KCl,
$\gamma_{\text{K}^{+}}>\gamma_{\text{Cl}^{-}}$, and for
$\Psi_{s}>0$, the contribution from the K$^{+}$ co-ions dominates over
that of the Cl$^{-}$ counter-ions, resulting in a net
increase of $\varepsilon_{s}$ as in Fig.~\ref{fig9}.
This phenomenon is more
prominent for higher $n_{b}$, because $\varepsilon_{s}(\Psi_{s}=0)$ is
further reduced, leading to a more substantial
permittivity increase due to the decrease in the K$^{+}$  co-ion concentration (Fig.~\ref{fig9}).

In summary, the non-monotonic dependence of $\varepsilon_{s}$ on
$\left|\Psi_s\right|$ is
expected to occur when the dielectric decrement of the co-ions is larger
than that of the counter-ions, $ \gamma_{{\rm K}^{+}}>\gamma_{{\rm Cl}^{-}}$.
At  high salt concentrations, when the PB model becomes invalid at any
surface potential,
this effect leads to a skewed bell-shape differential capacitance, and
the occurrence of the peak of bell-shaped $C$ at $\Psi_{s}\neq 0$.

%fig8
%%%%%%%%%%%%%%%%%%%%%%%%%%%%%%%%%%%%%%%%%%%%%%%%%%%%%%%%%%%%%%%%%%%%%%%%%%
%

%fig9
%%%%%%%%%%%%%%%%%%%%%%%%%%%%%%%%%%%%%%%%%%%%%%%%%%%%%%%%%%%%%%%%%%%
%

%%%%%%%%%%%%%%%%%%%%%%%%%%%%%%%%%%%%%%%%%%%%
\section{Effects of nonlinear decrement}
\label{sec:nonlinear_decrement}
%%%%%%%%%%%%%%%%%%%%%%%%%%%%%%%%%%%%%%%%%%%%

%%%%%%%%%%%%%%%%%%%%%%%%%%%%%%%%%%%%%%%%%%%%%%%%%%%%

The linear decrement model~\cite{Hasted1948Dielectric} is a good approximation for salt concentrations up to about,
$n_{b}\alt$\,2M.  Close to the surface,
where the counter-ion concentration is high, nonlinear dependence on the
counter-ion concentration is expected and the permittivity decrement usually
becomes weaker than for the linear decrement.
Because there are no direct experiments that reported the local variation of
$\varepsilon(n_\pm)$ inside the EDL,
we investigate the effect of nonlinear decrement by comparing several available models.

%%%%%%%%%%%%%%%%%%%%%%%%%%%%%%%%%
\subsection{Effect of quadratic term in $\varepsilon(n)$}
%%%%%%%%%%%%%%%%%%%%%%%%%%%%%%%%%%%%%%

%%%%NL%%%%%%%%%%%%%%%%%%%%%%%%%%%%%%%%%
To study the effect of nonlinear correction to the permittivity decrement,
we expand the permittivity up to second order in $n_{\pm}$
\begin{align}
\varepsilon(n_\pm) &=
 \varepsilon_{w}-\sum_{\alpha=\pm}\left(\gamma_{\alpha}n_{\alpha}(z) -\frac{1}{2}\beta_{\alpha}n_{\alpha}^{2}\right) \, ,
\label{eq:quadratic_decrement}
\end{align}
with $\gamma_{\pm}=-\partial \varepsilon/\partial n_{\pm}|_s$ as
before and $\beta_{\pm}=\partial^{2}\varepsilon/\partial
n_{\pm}^2|_s$, evaluated at the same $z=0$ surface.  Because the bulk permittivity, $\varepsilon(n_b)$,
is known from experiments to be
a concave function~\cite{Wei1990Dielectric,Wei1992Ion,levy12:_dielec_const_of_ionic_solut,Levy2013Dipolar},
we assume $\beta_{\pm}\geq 0$.

Analytical results can be obtained as before with the assumption that the co-ions are completely depleted from the
negatively charged surface, $n_s^{-}\to 0$. This is a very good approximation as long as $|\Psi_s|$ is not too small.
Using it we get
\begin{equation}
\label{eps_beta0}
\varepsilon_s \simeq \varepsilon_w - \gamma_{+}n_s^{+} +
\frac{1}{2}\beta_{+}(n_s^{+})^2 \, .
\end{equation}
We define (as before) a related surface permittivity $\Delta\varepsilon_s$ as:
\begin{eqnarray}
\Delta\varepsilon_s
&=&\varepsilon_s+n_s^{+}\left.\frac{\partial \varepsilon}{\partial n_{+}}\right|_{s}\nonumber\\
&=&\varepsilon_w - 2\gamma_{+}n_s^{+} +
\frac{3}{2}\beta_{+}(n_s^{+})^2 \nonumber\\
&=& \varepsilon_s - \gamma_{+}n_s^{+} + \beta_{+}(n_s^{+})^2\, .
\label{eps_beta}
\end{eqnarray}
From the osmotic pressure expression, Eq.~(\ref{P_os}), it is easy to derive the modified
Grahame equation~\cite{2006Soft,Israelachvili2011Intermolecular} as applied to the non-linear $\varepsilon(n_\pm)$:
\begin{align}
\sigma^{2} &\simeq
\frac{\varepsilon_{0}\left(\varepsilon_{s}\right)^{2}}{\Delta\varepsilon_s} \frac{2k_{B}T}{a^{3}}
\ln \left(\frac{\phi_b}{\phi_s}\right) \, .
\label{eq:pmpb_sigma_ns2a}
\end{align}
Note that it has exactly the same form as
Eq.~(\ref{mod_grahame}) (the linear decrement),
but with different $\varepsilon_{s}$ and $\Delta\varepsilon_{s}$
as in Eqs.~(\ref{eps_beta0})-(\ref{eps_beta}) for the non-linear case.

The differential capacitance $C$ also
depends on $\beta_{+}$. It is obtained as the parametric
function of $n_{s}^{+}$, (just as in Eq.~(\ref{C_parametric})),
where
\begin{align}
\Psi_{s}
&=\ln\left(\frac{n_{b}}{n_{s}^{+}}\right)\nonumber\\
&\nonumber\\
&-\frac{\Delta\varepsilon_{s}+
\left(\gamma_{+}-\beta_{+}n_{s}^{+}\right)a^{-3}}{\Delta\varepsilon_{s}}
\ln\left(\frac{\phi_{b}}{\phi_{s}}\right) \, .
\label{eq:psi_nsNL}
\end{align}
and
\begin{align}
\frac{\partial \Psi_{s}}{\partial n_{s}^{+}}
=&-\frac{1}{n_{s}^{+}}
\nonumber\\&\nonumber\\
&-\frac{\Delta\varepsilon_{s}+
\left(\gamma_{+}-\beta_{+}n_{s}^{+}\right)a^{-3}}
{\Delta\varepsilon_{s}}
\left(\frac{a^{3}}{\phi_{s}}\right)
\nonumber\\&\nonumber\\
&+\frac{\varepsilon_{s}\beta_{+}-2\left(\gamma_{+}-\beta_{+}n_{s}^{+}\right)^{2}}
{a^{3}(\Delta\varepsilon_{s})^{2}}
\ln\left(\frac{\phi_{b}}{\phi_{s}}\right)\, ,
\label{eq:dpsi_dnsNL}
\end{align}

\begin{align}
\frac{\partial\sigma}{\partial n_{s}^{+}}
=&\frac{k_{B}T}{\sigma\phi_s} \frac{\varepsilon_{0}\varepsilon_{s}^2}{\Delta\varepsilon_{s}}
\nonumber\\&\nonumber\\
& +\frac{\sigma n_{s}^{+}}{2\varepsilon_{s}\Delta\varepsilon_{s}}
\left[2\left(\gamma_{+}-\beta_{+}n_{s}^{+}\right)^{2}
-\varepsilon_{s}\beta_{+}\right]\,,
\label{eq:dsigma_dnsNL}
\end{align}

Nonlinear correction of the permittivity decrement at high counter-ion
concentration implies that the concentration at the dielectrophoretic
saturation should change, as is clearly seen from the modified Grahame
equation, Eq.~(\ref{eq:pmpb_sigma_ns2a}).  By solving
Eq.~(\ref{eps_beta}) with non-zero $\beta_{+}>0$, the
concentration at dielectrophoretic saturation becomes
\begin{align}
 n_{s}^{+} &=
\frac{2\gamma_{+}-2\sqrt{\gamma_{+}^{2}-3\beta_{+}\varepsilon_{w}/2}}{3\beta_{+}}
\nonumber\\
&\nonumber\\
&\simeq
\frac{\varepsilon_{w}}{2\gamma_{+}}
\left(1+\frac{3\beta_{+}\varepsilon_{w}}{8\gamma_{+}^{2}} \right) \, ,
\end{align}
where we assume that the quadratic term represents a small correction,
$0<\beta_{+}\ll 2\gamma_{+}^{2}/3\varepsilon_{w}$.  It shows that the
dielectrophoretic saturation concentration shifts to higher values of
$n_s^{+}$ as compared with those predicted by the linear decrement. The
effective size of the dielectrophoretic saturation for the nonlinear decrement (denoted as
$d^{\text{NL}}$) becomes smaller than that of the linear decrement,
$d^{\text{NL}}<d$.

\begin{figure}[htbp]
 \center
 \includegraphics{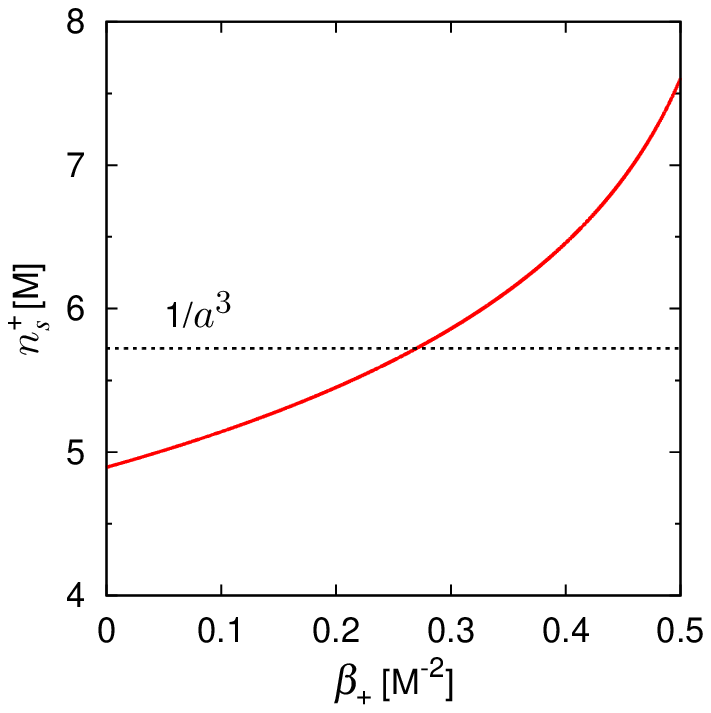}
\caption{\textsf{
The concentration at the dielectrophoretic saturation for the quadratic
 permittivity model, Eq.~(\ref{eps_beta0}), as a function of the
 quadratic coefficient, $\beta_{+}$.
The parameters are $\varepsilon_w=78.3$ and $\gamma_{+}=8$~(K$^{+}$ ions).
The dotted line indicates the closed-packing concentration for the
 K$^{+}$ ions.
}}
\label{fig10}
\end{figure}

In Fig.~\ref{fig10}, we plot $n_{s}^{+}$ at dielectrophoretic saturation
for K$^{+}$ ions, as a function of the quadratic coefficient,
$\beta_{+}$.  The figure shows that $n_{s}^{+}$ at the dielectrophoretic
saturation increases with $\beta_{+}$.  Moreover, when the quadratic
coefficient is considerably large, $\beta_{+}>
2\gamma_{+}^{2}/3\varepsilon_{w}$, the dielectrophoretic saturation does not occur, and
$d^{\text{NL}}\to 0$ in this extreme case.  The increase of the
dielectrophoretic saturation concentration implies a possible change of
the working saturation mechanism towards the sterically-dominant one.
For the K$^{+}$ ions, the dielectrophoretic saturation is predicted for the linear decrement model
because $d/a>1$ (see Table~\ref{tbl:parameters}).  This can also be
observed for $\beta_{+}=0$ in Fig.~\ref{fig10}.  The change of the
saturation mechanism occurs for $\beta_{+}\gtrsim 0.27$, where
$n_s^{+} \gtrsim 1/a^3$ and the condition $d^{\text{NL}}/a<1$ is
satisfied.

In general, weakening the permittivity decrement ($\beta^{+}>0$) at higher
ion  concentration increases the value of $n_s^+$ at the dielectrophoretic
saturation, and may even lead to disappearance of this saturation. The latter scenario
depends on the specific type of the nonlinear decrement model as will be demonstrated next.

%fig10
%%%%%%%%%%%%%%%%%%%%%%%%%%%%%%%%%%%%%%%%%%%%%%%%%%%%%%%%%%%%%%%%%%%%%%%%%%%%%%%%%%%%%%
%%%%%%%%%%%%%%%%%%%%%%%%%%%%%%%%%%%%%%%%%%%%%%%%%%%%%%%%%%%%%%%%%%%%%%%%%%%%%%%%%

%%%%%%%%%%%%%%%%%%%%%%%%%%%%%%%%%%%%%%%%%%%%%%%%%%%%%%%%%%%%%%%%%%
\subsection{Non-linear $\varepsilon(n)$ of a one-loop expansion }
%%%%%%%%%%%%%%%%%%%%%%%%%%%%%%%%%%%%%%%%%%%%%%%%%%%%%%%%%%%%%%%%%%%

An analytic expression for bulk permittivity of electrolyte solutions
has been derived
in Refs.~\onlinecite{levy12:_dielec_const_of_ionic_solut,Levy2013Dipolar}
using the dipolar Poisson--Boltzmann (DPB) model. The free energy of the electrolyte solution
takes into account the ionic and dipolar degrees of freedom. The field-theoretical
calculation employs
a one-loop expansion of the free energy beyond
mean-field theory, and accounts for the fluctuations in ion and dipolar solvent concentrations.
At high salinity, the permittivity function, $\varepsilon(n_b)$, has a nonlinear
dependence on the salt concentration $n_b$, and  fits nicely experiments that
also observed nonlinear behavior of $\varepsilon(n_b)$ in that high salt range.
The ionic solution permittivity $\varepsilon(n_b)$ was found to be:
\begin{align}
\label{subeq:lao_permittivity}
\varepsilon_{w} &=\varepsilon_{\rm DPB}
+\frac{\left(\varepsilon_{\rm DPB}-\varepsilon_{0}
\right)^{2}}{\varepsilon_{\rm DPB}}
\frac{4\pi}{3c_{d}b^{3}} \, ,
\nonumber\\
\varepsilon_{\rm DPB} &=\varepsilon_{0}
+\frac{p_{0}^{2}c_{d}}{3k_{B}T},
\nonumber\\
\varepsilon(n_{b}) &=\varepsilon_w\nonumber\\
&-\frac{\left(\varepsilon_{\rm DPB}-\varepsilon_{0}
\right)^{2}}{\varepsilon_{\rm DPB}}
\frac{\kappa_{D}^{2}}{\pi c_{d}b}
\left(1-\frac{\kappa_{D}b}{2\pi}\tan^{-1}
\frac{2\pi}{\kappa_{D}b} \right) \, ,
\end{align}
where $p_{0}=1.8$\,D is the water dipolar moment,
$c_{d}=55$\,M is the water molar density,
$\kappa_{D} = \sqrt{8\pi l_{B}n_{b} }$ is the inverse Debye-H\"uckel length,
$l_{B} ={e^{2}}/({4\pi \varepsilon_{\rm DPB}k_{B}T})$
is the Bjerrum length, $T=300$~K, and $b$ is the microscopic cutoff length.
The latter  incorporates into its value the distance of closest approach between all types of ions and dipoles, and
is used as a fitting parameter to experimental results.
Note that in the DPB model, $\varepsilon_w$ of Eq.~(\ref{subeq:lao_permittivity})
is a function of the water dipolar moment ($p_0$) and density ($c_d$).
The permittivity of the ionic solution, $\varepsilon(n_b)$, depends in addition
on the ions and their interactions with the dipoles.

Because we are interested in the EDL behavior, we need to extract the explicit
dependence of $\varepsilon$ on the local counter-ion and co-ion
concentrations, $n_{\pm}(z)$.
The permittivity of Eq.~(\ref{subeq:lao_permittivity}) depends on
$n_{b}$ through $\kappa_{D}$.
As it is difficult to separate the contributions from cations and anions in Eq.~(\ref{subeq:lao_permittivity}),
we replace $n_{b}$ with $(n_{+}+n_{-})/2$, and get as an approximation:
\begin{align}
\varepsilon(n_{b}) &\simeq \varepsilon \left(\frac{n_{+}+n_{-}}{2}\right) \, .
\label{eq:lao_ion}
\end{align}

\begin{figure}[htbp]
 \center
 \includegraphics{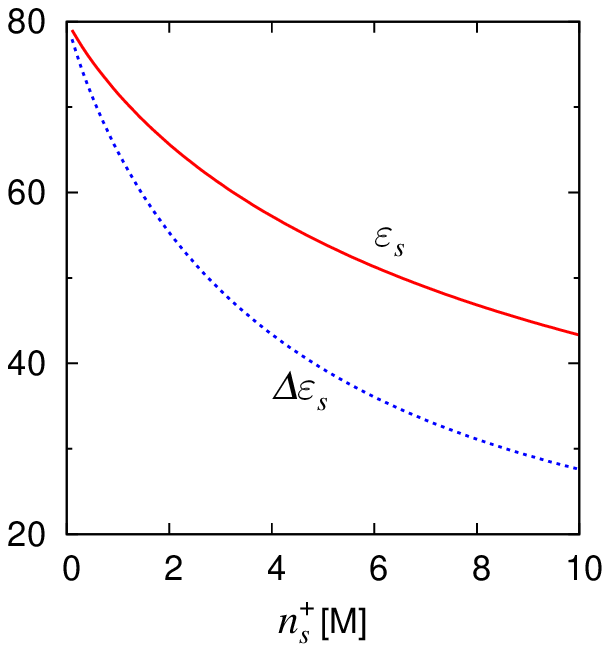}
\caption{\textsf{
The surface $\varepsilon_s$ of the
nonlinear permittivity model of Refs.~\onlinecite {levy12:_dielec_const_of_ionic_solut,Levy2013Dipolar}, as in
Eqs.~(\ref{subeq:lao_permittivity}) and (\ref{eq:lao_ion}) with $b=0.264$\,nm (solid red line),
and $\Delta\varepsilon_s$ (dashed blue line), Eq.~(\ref{delta_eps_DPB}),
as a function of the counter-ion concentration at the
surface, $n_s^{+}$. No dielectrophoretic saturation in the
 counter-ion concentration is seen as $\Delta\varepsilon_s>0$ for the entire range of $n_s^{+}$.
}}
\label{fig11}
\end{figure}

The function $\Delta\varepsilon_s$ for the dielectrophoretic saturation
condition
can be calculated analytically from  Eq.~(\ref{eq:lao_ion}) by setting $n_{s}^{-}=0$.
Namely, at the surface, $\varepsilon_{s}(n_{s}^+,n_s^{-}=0)$ and,
\begin{align}
\Delta\varepsilon_{s}(n_{s}^+)
=\varepsilon_{w}-\frac{
\left(\varepsilon_{\rm DPB}-\varepsilon_{0}\right)^{2}}{\varepsilon_{\rm DPB}^{2}}
\frac{n_{s}^{+} e^{2}}{ 2\pi k_{B}Tc_{d}b}
\nonumber\\
\times \left(4-5\frac{\kappa_{D}b}{2\pi}
\tan^{-1}\frac{2\pi}{\kappa_{D}b}+
\frac{1}{1+(\frac{2\pi}{\kappa_{D}b})^{2}}\right) \, .
\label{delta_eps_DPB}
\end{align}
From Eq.~(\ref{eq:dielectrophoretic_saturation_condition}), the
condition to satisfy for dielectrophoretic saturation is
$\Delta\varepsilon_{s}(n_{s}^+)=0$. However, this function in the DPB
model is concave and approaches a positive definite value,
$\Delta\varepsilon_s\to\varepsilon_{\rm DPB}$ as
$n_{s}^{+}\to\infty$. Hence, as can be observed from Fig.~\ref{fig11},
the $\Delta\varepsilon_{s}(n_{s}^+)$ expression is always positive, and
the dielectrophoretic saturation cannot occur for this permittivity
function.  In other words, for the Levy--Andelman--Orland permittivity
model, the counter-ion saturation at high
surface charge is always sterically-dominant.

%fig11
%%%%%%%%%%%%%%%%%%%%%%%%%%%%%%%%%%%%%%%%%%%%%%%%%%%%%%%%%%%%%%%%%%%%%%%%%%%%%%%%%%%%%%
%%%%%%%%%%%%%%%%%%%%%%%%%%%%%%%%%%%%%%%%%%%%%%%%%%%%%%%%%%%%%%%%%%%%%%%%%%%%%%%%%

\begin{figure}[htbp]
 \center
\includegraphics{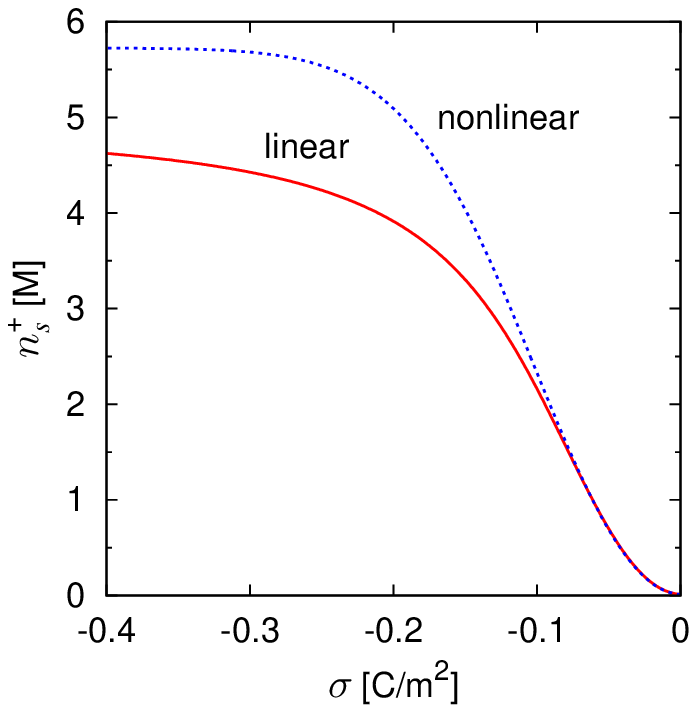}
\caption{ \textsf{
Counter-ion concentration, $n_{s}^+$,  in contact with a negatively
charged surface, as a function of the surface charge density,
$\sigma$, for KCl with $n_{b}=10$\,mM.  The dotted (blue) line corresponds to the nonlinear
permittivity decrement of Eqs.~(\ref{subeq:lao_permittivity}) and
(\ref{eq:lao_ion}) with $b=0.264$~nm, while the solid (red) line
corresponds to the linear permittivity decrement, Eq.~(\ref{eq:linear_decrement}).
}}
\label{fig12}
\end{figure}

Figure~\ref{fig12} shows a comparison
between the linear decrement model as shown in Sec.~\ref{sec:generalized_grahame}, and the nonlinear decrement model of
Eqs.~(\ref{subeq:lao_permittivity}) and (\ref{eq:lao_ion}).
The counter-ion concentration in contact with a negatively charged
surface, $n_{s}^{+}$, is plotted as a function of $\sigma$ for KCl with $n_{b}=10$\,mM.
For the linear decrement model, $n_{s}^{+}$ of the K$^+$ ions
approaches the concentration of the dielectrophoretic saturation
($n_s^+\simeq 4.89$\,M) because for K$^+$, $d/a\simeq 1.05 >1$,
while for the
nonlinear decrement model of
Refs.~\onlinecite{levy12:_dielec_const_of_ionic_solut,Levy2013Dipolar}, the
surface counter-ion concentration approaches the closed-packing
one of K$^{+}$, $n_s^+\simeq 5.72$\,M.

%fig12
%%%%%%%%%%%%%%%%%%%%%%%%%%%%%%%%%%%%%%%%%%%%%%%%%%%%%%%%%%%%%%%%%%%%

\begin{figure}[htbp]
 \center
\includegraphics{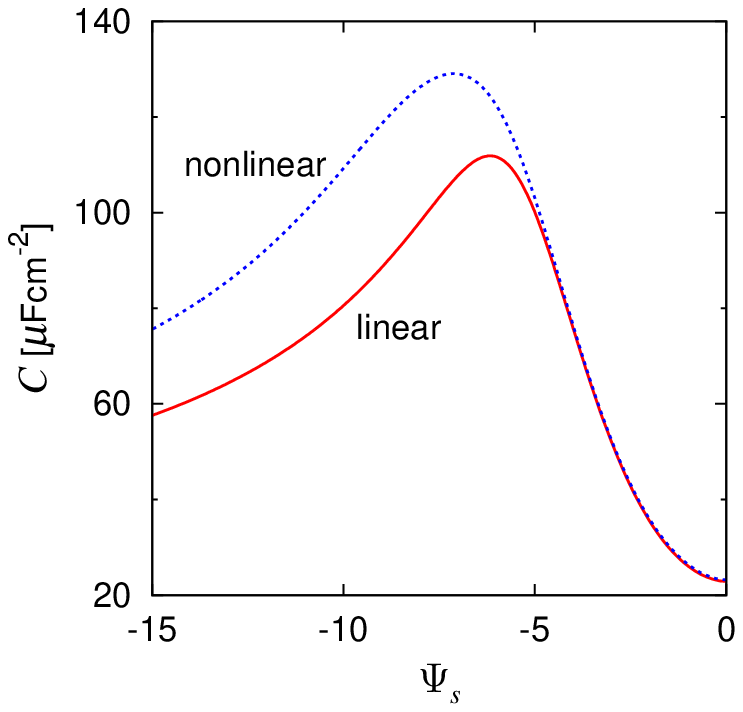}
\caption{\textsf{
Differential capacitance $C$ for
KCl with $n_{b}=10$\,mM, as function of negative $\Psi_s$.
The blue dotted
line denotes the nonlinear permittivity decrement of Eq.~(\ref{subeq:lao_permittivity})
with $b=0.264$\,nm, while the red solid line corresponds to a linear permittivity
decrement.
}}
\label{fig13}
\end{figure}

We conclude this section by comparing in Figure~\ref{fig13} the
differential capacitance, $C$, as obtained in
the linear
decrement model, Eq.~(\ref{eq:linear_decrement}), and the nonlinear decrement one, of
Eqs.~(\ref{subeq:lao_permittivity}) and (\ref{eq:lao_ion}). The calculations are done
as a
function of $\Psi_{s}<0$ for KCl with
$n_{b}=10$\,mM.
For the nonlinear decrement model, the peak position occurs at higher $\Psi_s$ than that for the linear decrement.
This reflects the fact
that the saturation of counter-ion concentration is dominated by steric effects.
Moreover, due to the weaker permittivity decrement in the nonlinear
model, the $C$ value at the peak is
higher than that for the linear decrement.
%%%%%%%%%%%%%%%%%%%%%%%%%%%%%%%%%%%%%%%%%%%%%%%%%%%%%%%%%%%%%%%%%%%%%%%%

%fig13
%%%%%%%%%%%%%%%%%%%%%%%%%%%%%%%%%%%%%%%%%%%%%%%%%%%%%%%%%%%%%%%%%%%%%%%
%%%%%%%%%%%%%%%%%%%%%%%%%%%%%%%%%%%%%%%%%%%%%%%%%%%%%%%%%%%%%%%%%%%%%%%%%%%%%%%%%%

%%%%%%%%%%%%%%%%%%%%%%%%%%%%%%%%%%%%%%
\section{Conclusions}
%%%%%%%%%%%%%%%%%%%%%%%%%%%%%%%%%%%%%%

We have presented  the combined effect of ion finite-size and
dielectric decrement on the electric double-layer (EDL) properties. The two important
parameters are the
surface charge density $\sigma$ (or equivalently surface potential, $\psi_s$), and the salt concentration,
$n_b$. Analytic expressions of the modified Grahame equation and  the differential
capacitance $C$ were derived, for several models of ionic-dependent permittivity,
$\varepsilon(n_{\pm})$.

We first treat the simpler linear decrement model,
$\varepsilon(n_\pm)=\varepsilon_w-\gamma_{+}n_{+}-\gamma_{-}n_{-}$.  The
counter-ion concentration at the surface proximity, $n_s$, exhibits a
saturation at high $\sigma$.  It originates either from the steric or
dielectric decrement effects.  Within the linear decrement model, the
dominant mechanism of the counter-ion saturation is ionic specific, and is determined by the relative size
of $d=(2\gamma/\varepsilon_w)^{1/3}$ and the ionic finite size $a$.  For
$d>a$, the dielectric decrement dominates and the counter-ion
concentration at the surface proximity saturates at the
dielectrophoretic saturation, $n_s\simeq \varepsilon_{w}/2\gamma$, which
then gives a lower and non-zero bound to the surface permittivity,
$\varepsilon_s\simeq \varepsilon_{w}/2$.  For $d<a$, although the
dominant saturation is the
steric one with $n_s\simeq 1/a^3$, the differential capacitance is found
to be strongly affected by the dielectric decrement.

At low salt concentrations, whether the dominant mechanism
for counter-ion saturation is
dielectrophoretic or steric, the differential capacitance $C$ exhibits a
camel-shape as a function of $\psi_{s}$. This is obtained by our
analytic and numerical results.
In contrast, at high salt concentrations, the
differential capacitance exhibits a skewed bell-shape, where the
uni-modal peak is shifted from $\psi_{s}=0$. This shift originates from
the asymmetry between the cation and anion polarization properties.

We also discuss possible effects of nonlinear permittivity decrement
on the dielectrophoretic saturation and
differential capacitance. When a nonlinear $\varepsilon(n)$
is considered, the
concentration at the dielectrophoretic saturation becomes larger than
that of the linear decrement. Therefore, the effective length parameter
$d$ for the non-linear model becomes smaller than for the linear
decrement case.  Moreover, the
dielectrophoretic saturation does not exist when  the
permittivity decrement is too weak.
In such a case, the peak in $C$ always originates from
the sterically-dominant saturation.

Finally, we would like to mention the possibility of direct experimental
determination of the ionic specific dielectric decrement.
In the past,  measurements of the bulk permittivity have shown that
$\varepsilon(n_b)$ depends on the salt concentration, $n_b$,
but it was not possible to
separate the contributions coming from the cations or anions.
In order to evaluate the separate contribution of each ion type,
appropriate physical quantity other than the bulk
permittivity is required.
The differential capacitance of the EDL at high surface
potentials is one such candidate, because it essentially depends
only on the dielectric decrement only by the counter-ions.
The analytic relationship between the counter-ion specific decrement and the
differential capacitance,
Eqs.~(\ref{eq:psi_nsNL})-(\ref{eq:dsigma_dnsNL}),
would give a way to evaluation directly the ionic-specific dielectric
decrement.
Because our analytic results are valid for general nonlinear permittivity
decrement,
they can be used to determine nonlinear permittivity behavior for high
ionic concentrations in contact with highly charged surfaces.

\bigskip

\noindent
%%%%%%%%%%%%%%%%%%%%%%%%%%%%%%
{\bf Acknowledgments}~~
%%%%%%%%%%%%%%%%%%%%%%%%%%
%
We thank M. Bazant, M. Biesheuvel, C. Grosse,
G. I. Guerrero-Garc\'{i}a and
M. Olvera de la Cruz
for useful discussions and numerous suggestions.
The numerical calculations have been partly carried out using the
computer facilities at the Research Institute for Information
Technology, Kyushu University.
Y.N. would like to acknowledge support
from the JSPS Institutional Program for Young Researcher Overseas
Visits and the hospitality of Tel Aviv University where this project was initiated.
This work has been supported by Grants-in-Aid for Scientific Research (JSPS
KAKENHI) under Grant No.~26400433,
the Israel Science Foundation~(ISF) under Grant No.~438/12,
and the U.S.-Israel Binational Science Foundation~(BSF) under Grant No.~2012/060.

\end{document}